\documentclass[manuscript,12pt]{aastex}

\title{
Planetary Growth with Collisional Fragmentation and Gas Drag
}

\author{Hiroshi Kobayashi$^{1}$, Hidekazu Tanaka$^2$, Alexander V.
Krivov$^1$, Satoshi Inaba$^3$}

\affil{$^{1}$ Astrophysical Institute and University Observatory,
Friedrich Schiller University, Schillergaesschen 2-3, 07745, Jena, GERMANY}

\affil{$^{2}$ Institute of Low Temperature Science, Hokkaido University,
Kita-Ku Kita 19 Nishi 8, Sapporo 060-0819, JAPAN}

\affil{$^{3}$ School of International Liberal Studies, Waseda University,
1-6-1 Nishi-Waseda, Shinjuku-ku, Tokyo 169-8050, JAPAN}

\email{hkobayas@astro.uni-jena.de}
\affil{}\affil{}\affil{}
\affil{}\affil{}\affil{}\affil{}\affil{}
\affil{}\affil{}\affil{}\affil{}\affil{}
\affil{
\begin{flushleft}
Manuscript pages: 53                   \\
Figures: 13                           \\
Tables: 1\\
\end{flushleft}
}
\affil{
\begin{flushleft}
Running head: Planetary Growth with Fragmentation and Gas Drag
\end{flushleft}
}
\affil{}\affil{}\affil{}\affil{}\affil{}\affil{}\affil{}\affil{}\affil{}\affil{}
\affil{
\begin{flushleft}
Correspondence Address:\\
Hiroshi Kobayashi \\
Astrophysical Institute and University Observatory
Friedrich Schiller University
Schillergaesschen 2-3, 07745 Jena, GERMANY\\ 
Tel: +49 3641 947 526\\
Fax: +49 3641 947 532\\
hkobayas@astro.uni-jena.de
\end{flushleft}
}
\affil{}\affil{}\affil{}

\begin{document}

\begin{abstract}
As planetary embryos grow, gravitational stirring of planetesimals by
embryos strongly enhances random velocities of planetesimals and makes
collisions between planetesimals destructive. The resulting fragments
are ground down by successive collisions.  Eventually the smallest
fragments are removed by the inward drift due to gas drag.  Therefore,
the collisional disruption depletes the planetesimal disk and inhibits
embryo growth.  We provide analytical formulae for the final masses of
planetary embryos, taking into account planetesimal depletion due to
collisional disruption.  Furthermore, we perform the statistical
simulations for embryo growth (which excellently reproduce results of
direct $N$-body simulations if disruption is neglected).  These
analytical formulae are consistent with the outcome of our statistical
simulations.  Our results indicate that the final embryo mass at several
AU in the minimum-mass solar nebula can reach about $\sim 0.1$ Earth
mass within $10^7$ years.  This brings another difficulty in formation
of gas giant planets, which requires cores with $\sim 10$ Earth masses
for gas accretion. However, if the nebular disk is 10 times more massive
than the minimum-mass solar nebula and the initial planetesimal size is
larger than 100 km, as suggested by some models of planetesimal
formation, the final embryo mass reaches about 10 Earth masses at
3--4\,AU.  The enhancement of embryos' collisional cross sections by
their atmosphere could further increase their final mass to form gas
giant planets at $5$--$10$\,AU in the solar system.

\end{abstract}


\section{INTRODUCTION}

In the standard scenario of planetary formation, terrestrial planets and
cores of Jovian planets are formed through the accretion of
planetesimals with initial size of 10--100\,km (e.g., Hayashi et
al. 1985).  This process called planetary accretion has been
investigated by statistical simulations (e.g., Wetherill and Stewart
1993; Inaba et al. 2003; Kenyon and Bromley 2004), by $N$-body
simulations (e.g., Kokubo and Ida 1996; 2002), and by the hybrid method
that combines an $N$-body simulation for large bodies called planetary
embryos with a statistical simulation for small bodies (e.g., Bromley
and Kenyon 2006; Chambers 2008).  Assuming that each collision of bodies
leads to a perfect agglomeration, the growth of planetesimals can be
accurately computed with $N$-body codes.  Inaba et al. (2001) showed
that the evolution of mass distribution and velocity dispersion of
bodies calculated by $N$-body simulation were reproduced by the
statistical simulation applying the collision rate and the
velocity-dispersion-evolution rate based on the results of orbital
integrations. Therefore, statistical simulations are reliable unless
planetary embryos collide with each other after developing a long-term
orbital instability.

Planetary embryos initially form through the runaway growth (e.g.,
Wetherill and Stewart 1989; Kokubo and Ida 1996).  The embryos keep
their orbital separations and grow through collisions with surrounding
planetesimals (Kokubo and Ida 1998).  At this stage, planetesimals of
almost initial size dominate the total mass of bodies (surface density).
When embryos reach about the mass of Mars, the velocity dispersion of
planetesimals is increased by the gravitational scattering at the
embryos.  Accordingly, another growth regime sets in, referred to as
oligarchic growth.  At this stage, higher velocities of planetesimals
cause their collisional fragmentation.  After a chain of successive
destructive collisions, often called ``collision cascade'', bodies get
smaller and smaller until they are removed by gas drag in protoplanetary
disks or by radiation pressure and/or Poynting-Robertson drag in debris
disks.  As a result, collision cascade decreases the surface density,
which slows down the growth of planetary embryos (Inaba et al. 2003;
Kenyon and Bromley 2008).

The planetary core (embryo) exceeding the critical core mass as large as
10 Earth masses can no longer retain a hydrostatic envelope, resulting
in the gas accretion and formation of the gas giant planets (e.g.,
Mizuno et al.\ 1980; Bodenheimer \& Pollack 1986; Pollack et al.\ 1996;
Ikoma et al.\ 2001).  However, since the embryo growth is hindered by
the planetesimal depletion in collision cascade, the final embryo can
hardly grow beyond the Mars mass and thus cannot form a gas giant
(Kobayashi and Tanaka 2010).

However, it is possible that fragmentation inhibits the embryo growth to
a lesser extent than ascertained before.  Collisional fragments orbiting
a central star drift inward by gas drag.  The drift time shortens as
bodies become smaller by collision cascade, until at a certain size
their motion gets coupled with gas motion.  Such coupled bodies have
lower drift velocity and thus can survive around planetary embryos for
longer time.  Kenyon and Bromley (2009) suggested that accretion of
those coupled bodies by planetary embryos may promote further growth of
the latter.  In their simulations, they assumed that the coupled bodies
no longer experience an inward drift.  However, the drift, although at a
reduced rate, is important to determine the embryo mass gain due to the
accretion of coupled bodies. 

This paper investigates the embryo growth taking into account the
accretion of fragments resulting from the collision cascade before their
removal by gas drag.  To this end, we perform an analytic study that
extends the model of Kobayashi and Tanaka (2010) by including the
removal by gas drag, as well as statistical simulations.  In the
numerical treatment, we do not neglect the drift of coupled bodies by
gas drag.  Instead, we take into account that the gas drag law changes
for small bodies in the both analytical and numerical procedure (e.g.,
Adachi et al. 1976).  The goal is to find out to what extent the
collisional fragmentation combined with gas drag would affect the embryo
growth and whether an embryo can reach the critical core mass.

We develop the analytic theory and derive the final embryo mass with
fragmentation and gas drag in Section 2.  In Section 3, we check the
formulae for the final mass against the statistical numerical
simulations. Section 4 contains a summary and a discussion of our
results.

\section{THEORETICAL MODEL}
\label{sec:theo}

In this section, we summarize the oligarchic growth of planetary embryos
and the surface density decline resulting from fragmentation.  Then we
derive the general formulae for the final mass through the oligarchic
growth with fragmentation.

\subsection{DISK AND FRAGMENTATION MODEL}

We introduce a power-law disk model for the surface mass density of
solids and gas.  The solid surface mass density is taken to be 
\begin{equation}
 \Sigma_{\rm s,0} = f_{\rm ice} \Sigma_1 \left(\frac{a}{1{\rm AU}}\right)^{-q}
  \, {\rm g\, cm}^{-2}, 
\end{equation}
where $a$ is the distance from a central star, $\Sigma_1$ is the
reference surface density at 1\,AU, $q$ is the power-law index of
the radial distribution, and $f_{\rm ice}$ is the factor that
represents the increase of solid density by ice condensation beyond the snow line
$a_{\rm ice}$ at which the temperature equals the ice condensation
temperature $\simeq 170$\,K. We set the gas surface density to
\begin{equation}
\Sigma_{\rm gas,0} = f_{\rm gas}  \Sigma_1 \left(\frac{a}{1{\rm
					  AU}}\right)^{-q} 
  \, {\rm g\, cm}^{-2}, 
\end{equation}
where $f_{\rm gas}$ is the gas-dust ratio. In the nominal case, we adopt
$f_{\rm gas} = 240$ (Hayashi 1981). 
If the disk is optically thin, 
the gas temperature is given by 
\begin{equation}
 T = 280 \Biggl(\frac{a}{1\,{\rm AU}}\Biggl)^{-1/2}
  \left(\frac{L_*}{L_\sun}\right)^{1/4} \,{\rm K}, 
\end{equation}
where $L_*$ and $L_\sun$ are luminosities of the central star and the sun,
respectively. 
For $L_* = L_\sun$, this yields $a_{\rm ice} = 2.7\,$AU. 
In the minimum-mass solar nebula (MMSN) model, $\Sigma_1 = 7 \,{\rm g/cm}^2$,
$q = 3/2$ and $f_{\rm ice} = 1$ ($a<a_{\rm ice}$) and 4.2 ($a>a_{\rm
ice}$). 
However, since a large amount of small dust is present even after
completion of planetesimal formation, the disk is expected to be
optically thick. This would make the temperature lower and its radial profile
different (e.g., Kusaka et al. 1970). However, these effects would not
drastically influence the embryo growth except for the location of the snow line. 

We take the fragmentation model described by Kobayashi and Tanaka (2010). 
Assuming that the fragmentation is scaled by the energy, 
the total ejecta mass $m_{\rm e}$ produced by one collision between
$m_1$ and $m_2$ is given by 
\begin{equation}
 \frac{m_{\rm e}}{m_1+m_2} = \frac{\phi}{1+\phi}.\label{eq:fragmass}  
\end{equation} 
Here the scaled impact energy $\phi$ is given by $ m_1 m_2 v^2 / 2
(m_1+m_2)^2 Q_{\rm D}^*$, where $v$ is the collisional velocity between
$m_1$ and $m_2$ and $Q_{\rm D}^*$ is the critical specific impact energy
needed to disrupt the colliding bodies and eject 50\% of their mass
($m_{\rm e} = (m_1 + m_2)/2$). The value of $Q_{\rm D}^*$ is given by
the larger of the two colliders ($m_1$ and $m_2$). Note that, since
Kobayashi and Tanaka (2010) separately determine the ejecta mass from
$m_1$ and $m_2$ in their analysis, Eq.~(\ref{eq:fragmass}) is different
from their definition.

The energy threshold is given by 
\begin{eqnarray}
\displaystyle
 Q_{\rm D}^* = Q_{\rm 0s} \Biggl(\frac{r}{1\,{\rm cm}}\Biggr)^{\beta_{\rm s}} +
Q_{\rm 0g} \rho \Biggl(\frac{r}{1\,{\rm cm}}\Biggr)^{\beta_{\rm g}} +
C_{\rm gg}
\frac{2Gm}{r}
,\label{eq:qd} 
\end{eqnarray}
where $r$ and $m$ are the radius and mass of a body and 
$\rho$ is its density. 
Benz and Ashpaug (1999) provide $Q_{\rm 0s}, {\beta_{\rm s}}, 
Q_{\rm 0g}$, and ${\beta_{\rm g}}$ for $r = 1$--$10^7\,$cm from the
hydrodynamical simulation of the collisional dispersion. 
The first term in the right-hand side of Eq.~(\ref{eq:qd})
controls $Q_{\rm D}^*$ for $r \la 10^4$--$10^5$\,cm and 
the second term describes $Q_{\rm D}^*$ of the larger bodies. 
For $r \ga 10^7$\,cm, $Q_{\rm D}^*$ is purely determined by the
gravitational binding energy, being independent of material properties. 
The collisional simulation for the gravitational aggregates yields
$C_{\rm gg} \simeq 10$ (Stewart and Leinhardt 2009).

\subsection{ISOLATION MASS}

Planetary embryos no longer grow after they have accreted all planetesimals
within their feeding zones. The width of a feeding zone is equal
to the orbital separation of the embryos,
$\tilde b (2M/3M_*)^{1/3} a$, where 
$M$ is the embryo mass, $M_*$ is the mass of central star, 
$\tilde b \simeq 10$ is a factor (Kokubo and Ida 2002).
Therefore, the maximum mass or ``isolation mass'' satisfy
$M_{\rm iso} = 2 \pi a^2  (2M_{\rm iso}/3M_*)^{1/3}\tilde b \Sigma_{\rm s}$.
It can be expressed as
\begin{equation}
 M_{\rm iso} = 2.8 \Biggl(\frac{\tilde b}{10}\Biggr)^{3/2}
  \Biggl(\frac{\Sigma_{\rm s,0}}{2.7\,{\rm g/cm}^2}\Biggr)^{3/2}
  \Biggl(\frac{a}{5\,{\rm AU}}\Biggr)^3
  \Biggl(\frac{M_*}{M_\sun}\Biggr)^{-1/2} M_{\oplus},\label{eq:M_iso} 
\end{equation}
where 
$M_{\oplus}$ is the Earth mass and $M_\sun$ is the solar mass. 
The planetary embryo mass approaches the isolation mass if
fragmentation is ignored (Kokubo and Ida 2000; 2002). 
However, if fragmentation is included, the final embryo mass is expected to
be smaller.

\subsection{PLANETARY GROWTH}

\subsubsection{GROWTH WITHOUT ACCRETION OF FRAGMENTS}

At the runaway stage of planetary growth, the larger planetesimals grow
faster than smaller ones.  The solid surface density at the runaway
growth stage is determined by relatively small planetesimals.  At the
later, oligarchic stage, planetary embryos become massive, start to
gravitationally stir up the planetesimals and induce their collisional
disruption.  The resulting fragments are quickly removed by the inward
drift due to gas drag. Thus the fragmentation would reduce the surface
density and hence the final embryo mass.  Here, taking into account the
depletion of planetesimals due to the fragmentation, we estimate the
final embryo mass.  We neglect the accretion of fragments onto the
embryos, whereas the accretion of fragments will be considered in
Section 2.3.2.

At the oligarchic stage, planetary embryos are distributed uniformly due
to their mutual gravitational interaction.  Since they cannot collide
with each other due to their large orbital separation, they grow slowly.
If a planetary embryo with mass $M$ collides with planetesimals of mass
$m$, its growth rate is given by
\begin{eqnarray}
 \frac{dM}{dt} &=& \int dm \,
  m n_{\rm s} a^2 h_{m,M}^2 \langle P_{\rm col} \rangle \Omega_{\rm K},\label{eq:dM_org} 
\end{eqnarray}
where $n_{\rm s}dm$ is the surface number density of planetesimals with
mass in the range of $[m,m+dm]$, $\Omega_{\rm K}$ is the Kepler angular
velocity, $h_{m,M} = [(m+M)/3M_*]^{1/3}$ is the dimensionless 
reduced Hill radius between $m$ and $M$, and $\langle P_{\rm col}
\rangle$ is the dimensionless collision rate.  We assume $m \ll M$, and
Eq.~(\ref{eq:dM_org}) reduces to
\begin{equation}
 \frac{dM}{dt} = C_{\rm acc} \Sigma_{\rm s} a^2 h_{\rm M}^2 
  \langle P_{\rm col} \rangle \Omega_{\rm K}, 
\label{eq:dm}
\end{equation}
where $\Sigma_{\rm s}$ is the surface density of planetesimals, $h_{M} =
(M/3M_*)^{1/3}$, and $C_{\rm acc}$ is the correction factor on the order
of unity.  Because the eccentricity dispersion $e^*$ and inclination
dispersion $i^*$ of planetesimals are much larger than $h_{M}$ for
kilometer-sized or larger planetesimals, the dimensionless collision
rate $\langle P_{\rm col} \rangle$ is given by (e.g., Inaba et al. 2001)
\begin{equation}
 \langle P_{\rm col} \rangle = \frac{C_{\rm col} {\tilde R} h_{M}^2}{
  e^{*2}}, \label{eq:pcol_high} 
\end{equation} 
where $C_{\rm col} \simeq 36$ for $e^* = 2i^*$.  Here, ${\tilde R} = (9
M_*/ 4 \pi \rho)^{1/3}/a$ is the embryo radius scaled by $h_M a$ and
independent of the embryo mass.  In this paper, we do not take into
account the enhancement of $\langle P_{\rm col} \rangle$ due to
atmosphere which an embryo is expected to acquire.  Such an enhancement
may be efficient for $M \ga 0.1 M_{\oplus}$ (Inaba and Ikoma 2003).
Note that embryos grow not only through collisions with swarm
planetesimals but also with neighboring embryos because their separation
scaled by the Hill radius decreases by their growth.  Furthermore,
taking into account mass distribution of planetesimals, the actual
accretion rate becomes larger.  In this paper, we apply $C_{\rm acc}
\simeq 1.5$.

Since $dM/dt \propto e^{*-2}$, we need to describe evolution of $e^*$.
Two mechanisms that affect $e^*$ are viscous stirring and damping by
gas.  The viscous stirring of planetary embryos raises the random
velocities of planetesimals. The stirring rate of $e^{*2}$ is given by
\begin{equation}
 \Biggl(\frac{de^{*2}}{dt}\Biggr)_{\rm VS} = n_{M} a^2 h_M^4 \langle P_{\rm VS} \rangle \Omega_{\rm K}.
\label{eq:stiring} 
\end{equation}
Since embryos are uniformly distributed with a separation $2^{1/3}
h_{\rm M} a {\tilde b}$, their surface number density $n_M$ is given by
\begin{equation}
 n_M = \frac{1}{2^{4/3} \pi {\tilde b} h_M a^2}.\label{eq:nM}
\end{equation}
For $e^* = 2 i^* \gg h_{M}$, the dimensionless stirring rate $\langle
P_{\rm VS} \rangle$ is given by (Ohtsuki et al. 2002)
\begin{equation}
 \langle P_{\rm VS} \rangle = \frac{C_{\rm VS} h_{M}^2}{e^{*2}} \ln (\Lambda^2 +1),\label{eq:pvs_high} 
\end{equation}
where $C_{\rm VS} \simeq 40$.  Although $\ln (\Lambda^2+1)$ weakly
depends on $e^*$, we assume $\ln (\Lambda^2+1) \simeq 3$ for this
analysis.  On the other hand, $e^*$ is damped by gas drag and the
damping rate is given by
\begin{equation}
 \Biggl(\frac{d e^{*2}}{dt}\Biggr)_{\rm gas} = - \frac{C_{\rm gas}}{\tau} e^{*3},\label{eq_de_gas} 
\end{equation}
where $C_{\rm gas} \simeq 2.1$ (e.g., Inaba et al. 2001) and 
\begin{equation}
 \tau = \frac{m}{\pi r^2 C_{\rm D} \rho_{\rm gas} v_{\rm K}}.\label{eq:tau_gas} 
\end{equation}
Here, $\rho_{\rm gas}$ is the gas density in the midplane, $v_{\rm K}$
is the Keplerian velocity, and $C_{\rm D}$ is the dimensionless gas drag
coefficient.  The latter is defined as a factor that appears in the
expression for the gas drag force acting on a planetesimal with radius
$r$: $C_{\rm D} \pi r^2 \rho_{\rm gas} u^2/2$ with $u$ being the
relative velocity between the planetesimal and gas. For kilometer-sized
or larger planetesimals, $C_{\rm D}$ is a constant (Adachi et
al. 1976). We assume that $e^*$ is determined by the equilibrium between
the stirring and the damping (Eqs.~(\ref{eq:stiring}) and
(\ref{eq_de_gas})).  Equating the stirring and damping rates, we obtain
\begin{equation}
 \frac{e^*}{h_{M}} = \left[
\frac{C_{\rm VS} \ln (\Lambda^2+1) \Omega_{\rm K}
  \tau}{2^{4/3}\pi {\tilde b} C_{\rm gas}}
  \right]^{1/5}.\label{eq:e_gas} 
\end{equation}
We can estimate $e^* \sim (\tau \Omega_{\rm K})^{1/5} h_M$.  Since $\tau
\Omega_{\rm K} \gg 1$ for the planetesimals (Adachi et al. 1976), our
assumption of $e^* \gg h_M$ for the derivation is valid.  Inserting
Eqs.~(\ref{eq:pcol_high}) and (\ref{eq:e_gas}) to Eq.~(\ref{eq:dm}), the
growth rate is found to be
\begin{equation}
 \frac{dM}{dt} = C_{\rm acc} \Sigma_{\rm s} a^2 C_{\rm col} {\tilde R} \Omega_{\rm K} 
  \left[
\frac{2^{4/3}\pi {\tilde b} C_{\rm gas}}{C_{\rm VS} \ln (\Lambda^2+1)
\Omega_{\rm K} \tau} 
  \right]^{2/5} h_{M}^2.\label{eq:dm_gas}   
\end{equation}
The time evolution of a planetary embryo in the oligarchic
growth mode, neglecting fragmentation, is summarized in Appendix A.

As embryos grow, the random velocity of planetesimals increases, making
collisions between planetesimals destructive.  Fragments produced by the
collisions get smaller and smaller by successive collisions (collision
cascade) until the smallest ones are brought inward by gas drag and are
lost to the central star.  Planetesimals with mass $m$ dominate the
surface density $\Sigma_{\rm s}$ during the oligarchic growth.  Thus,
the collision cascade induced by embryo growth reduces the surface
density.  The evolution of $\Sigma_{\rm s}$ due to a collision cascade
is given by (see Kobayashi and Tanaka 2010; Appendix B for a derivation)
\begin{equation}
\frac{d\Sigma_{\rm s}}{dt} = - \frac{\Sigma_{\rm s}^2 (2-\alpha)^2 h_0 \Omega_{\rm
  K}}{m^{1/3}} \left( \frac{v(m)^2}{2 Q_{\rm
				D}^*(m)}\right)^{\alpha-1} s_{123}(\alpha), 
\label{eq:sigma_dep_collision_cascade}
\end{equation}
where $h_0 = 1.1 \rho^{-2/3}$ and
\begin{equation}
s_{123}(\alpha) = 
\int_{0}^{\infty} d\phi 
\frac{\phi^{-\alpha}}{1+\phi} 
\left\{
\left[-\ln \frac{\epsilon \phi}{(1+\phi)^2} + \frac{1}{2-b} \right]\phi
+ \ln(1+\phi)
\right\}, 
\end{equation}
where $b$ is the power-law exponent of the mass distribution of ejecta
yielded by one collision between $m_1$ and $m_2$ and $\epsilon$ is a
factor in the expression for the mass of the largest ejecta: $m_{{\rm
L}} = \epsilon (m_1+m_2) \phi /(1+\phi)^2$ with a use of the scaled
impact energy $\phi$.  For a collision cascade, the mass distribution
exponent $\alpha$ of fragments is given by $\alpha = (11+3p)/(6+3p)$ for
$v(m)^2/Q_{\rm D}^*(m) \propto m^{-p}$ and $s_{123}$ is insensitive to
constants $\epsilon < 1$ and $b =1$--2 (Kobayashi and Tanaka 2010).  We
set $b = 5/3$ and $\epsilon = 0.2$. 

Dividing Eq.~(\ref{eq:dm_gas}) by
Eq.~(\ref{eq:sigma_dep_collision_cascade}) and integrating, we obtain
the final embryo mass,
\begin{eqnarray}
 M_{\rm c} &=& \left(\frac{2\alpha-1}{3}\right)^{3/(2\alpha-1)}
\left(\frac{a^2 C_{\rm acc} C_{\rm col} {\tilde R} m^{1/3}}{(2-\alpha)^2 h_0 s_{123}}\right)^{3/(2\alpha-1)}
  \left(\frac{v_{\rm K}^2}{2 Q_{\rm D}^*}\right)^{3(1-\alpha)/(2\alpha-1)}
\nonumber 
\\&&
\times 
  \left(3 M_*\right)^{-(4-2\alpha)/(2\alpha-1)} 
\left[
\ln \left(\frac{\Sigma_{\rm s,0}}{\Sigma_{\rm s}}\right)
\right]^{3/(2\alpha-1)}
\left[
\frac{C_{\rm VS} \ln (\Lambda^2+1) \Omega_{\rm K}
  \tau}{2^{4/3}\pi {\tilde b} C_{\rm gas}}
  \right]^{-6\alpha/5(2\alpha-1)}, 
\label{eq:m_frag}
\end{eqnarray}
where $v = e^* v_{\rm K}$.  This equation is valid unless the surface
density is much smaller than that of embryos, $\Sigma_{\rm s} \ll M_{\rm
c} n_{M}$.  Therefore, an estimate of $\Sigma_{\rm s}$ for the final
embryo mass is given by
\begin{equation}
 \frac{\Sigma_{\rm s}}{\Sigma_{\rm s,0}} = 
  \frac{C_{\Sigma_{\rm s}} M_{\rm c}}{2^{4/3} \pi {\tilde b} a^2 h_{M_{\rm c}}\Sigma_{\rm s,0}} 
  = C_{\Sigma_{\rm s}} \Biggl(\frac{M_{\rm c}}{M_{\rm iso}}\Biggr)^{2/3},\label{eq:ratio_sigma} 
\end{equation}
where $C_{\rm \Sigma_{\rm s}} \ll 1 $ is a constant.  Generally, $M_{\rm
c}$ should be smaller than $M_{\rm iso}$.

We assume $Q_{\rm D}^* = Q_{\rm 0g} \rho r^{\beta_{\rm g}}$ for the
gravity regime and thus $\alpha = [11+3(\beta_{\rm g} -
2/15)]/[6+3(\beta-2/15)]$.  For the minimum-mass solar nebula ($\Sigma_0
= 7\,{\rm g/cm}^2$ and $q=3/2$), Eq.~(\ref{eq:m_frag}) can be re-written
as
\begin{eqnarray}
 M_{\rm c} &=& 0.10
  \Biggl(\frac{\ln (\Sigma_{\rm s,0}/\Sigma_{\rm s})}{4.5}\Biggr)^{1.21}
  \Biggl(\frac{a}{5\,{\rm AU}}\Biggr)^{0.63}
  \Biggl(\frac{m}{4.2\times 10^{20}\,{\rm g}}\Biggr)^{0.48}
  \nonumber
  \\
 &&\times 
  \Biggl(\frac{M_*}{M_\sun}\Biggr)^{-0.28}
  \Biggl(\frac{Q_{\rm 0g}}{2.1\,{\rm erg\, cm}^3 / {\rm
  g}^2}\Biggr)^{0.89}
  \Biggl(\frac{\rho}{1\,{\rm g/cm}^3}\Biggr)^{1.85} M_{\oplus},\label{eq:M_c_est} 
\end{eqnarray}
where $\beta_{\rm g} = 1.19$ for ice.  Here, we estimate $\ln
(\Sigma_{s}/\Sigma_{\rm s,0}) \simeq 4.5$ from
Eq.~(\ref{eq:ratio_sigma}) with the use of $C_{\Sigma_{\rm s}} = 0.1$
for $M_{\rm c}=0.10 M_\oplus$ and $M_{\rm iso}= 2.8 M_{\oplus}$.  With
fragmentation, the final embryo mass given by Eq.~(\ref{eq:M_c_est})
becomes much smaller than the isolation mass, Eq.~(\ref{eq:M_iso}).  As
we will show in Section \ref{eq:num} with the aid of statistical
simulations, the planetesimal mass $m$ is about $100\,m_0$, where $m_0$
is the initial planetesimal mass.  We assume $m=100\,m_0$ and will
compare $M_{\rm c}$ with the embryo mass obtained through the
statistical simulation.

\subsubsection{GROWTH WITH ACCRETION OF FRAGMENTS}

For the steady-state mass distribution of collision cascade (see
Appendix A), the surface density of small fragments is much lower than
that of planetesimals and the accretion of fragments is insignificant
for embryo growth.  The collision cascade may terminate at a certain
mass where destructive collisions no longer occur due to low collisional
velocities damped by a strong gas drag.  The steady-state mass
distribution is achieved if fragments at the low-mass end of the cascade
are quickly removed by the gas drag.  However, when the $\Sigma_{\rm
s}$-decay time resulting from the collision cascade is shorter than the
removal time of the fragments by gas drag, the termination of the
collision cascade yields a large amount of fragments at the low-mass end
of the cascade and would dominate the solid surface density.  Here, we
examine the termination of the collision cascade and estimate the embryo
growth through accretion of such fragments.

For small fragments, the $e^*$-damping rate resulting from the gas drag
is given by (Adachi et al. 1976)
\begin{equation}
 \left(\frac{de^*}{dt}\right)_{\rm gas,f} = - \frac{e^* \eta}{\tau},\label{eq:de_gas_stokes}
\end{equation}
where $\eta = (v_{\rm K} - v_{\rm gas})/v_{\rm K}$ is the deviation of
the gas rotation velocity $v_{\rm gas}$ from the Keplerian velocity. For
$q = 3/2$, $\eta = 1.8\times 10^{-3} (a/1\,{\rm AU})^{1/2}$.  For small
fragments, $C_{\rm D}$ depends on $u$ and $e^*$ becomes much smaller
than $\eta$ by the gas drag.  Hence, the damping rate given by
Eq.~(\ref{eq:de_gas_stokes}) is different from Eq.~(\ref{eq_de_gas}).
Since the gas drag substantially damps $e^*$ of small fragments,
$\langle P_{\rm VS}\rangle = 73$ is independent of $e^*$ for $e^* \ll
h_M$.  Thus, the equilibrium condition between the stirring
(Eqs.~(\ref{eq:stiring})--(\ref{eq:nM})) and the damping
(Eq.~(\ref{eq:de_gas_stokes})) gives
\begin{equation}
 e^{*2} = \frac{h_M^3 \langle P_{\rm VS} \rangle \tau \Omega_{\rm K}}{2^{4/3} \pi {\tilde b}
  \eta}.\label{eq:e_small}
\end{equation}
Smaller fragments have low $e^*$ because of low $\tau$. 

The collision cascade will no longer operate for such low $e^*$.  The
collisional energy between bodies of mass $m$ is estimated to be $m
e^{*2} v_{\rm K}^2/4$ and is much smaller than energy threshold $2 m
Q_{\rm D}^*$ for $m_{\rm e} = m$ at the low-mass end of the cascade.
Therefore, at the low-mass end
\begin{equation}
 e^{*2} v_{\rm K}^2 = C_{\rm L} Q_{\rm D}^*, \label{eq:tau_Q} 
\end{equation}
where $C_{\rm L}$ is a constant. As we will show, the surface density of
fragments becomes higher at the low-mass end of the collision
cascade. Our simulation yields $C_{\rm L} \simeq 1$ at the low-mass end.

The surface density of the fragments decreases as a result of the radial
drift by gas drag.  From Eqs.~(\ref{eq:e_small}) and (\ref{eq:tau_Q}),
we estimate the scaled stopping time ${\tilde \tau}_{\rm stop}$ of the
fragments at the low-mass end of the cascade to be
\begin{eqnarray}
 {\tilde \tau}_{\rm stop} &=& \frac{\tau \Omega_{\rm K}}{\eta} = \frac{2^{4/3}\pi \tilde b}{h_M^3
  \langle P_{\rm VS}\rangle} \frac{C_{\rm L} Q_{\rm D}^*}{v_{\rm K}^2}\label{eq:tstop_est}
 \\
 &\simeq& 19\, C_{\rm L} \Biggl(\frac{a}{5\,{\rm AU}}\Biggr) 
\Biggl(\frac{M}{0.1  M_\oplus}\Biggr)^{-1}
  \Biggl(\frac{Q_{\rm D}^*}{3.1\times 10^6{\rm \,erg/g}}\Biggr), 
\end{eqnarray}
where $Q_{\rm D}^* \simeq 3.1\times 10^6 \,{\rm erg/g}$ for ice bodies
with $r = 10\,$m and we will derive $M \sim 0.1 M_\oplus$ in the
following analysis.  Since ${\tilde \tau}_{\rm stop} \gg 1$, the
coupling between the fragments and gas is weak.  Therefore, the drift
velocity of fragments is given by $2\eta^2 a /\tau$ (Adachi et
al. 1976).  Since $\tau \propto a^{3/4}$ for the fragments and
$\Sigma_{\rm s} \propto a^{-q}$, the $\Sigma_{\rm s}$-depletion rate is
given by $d\Sigma_{\rm s}/dt = - 2 (9/4-q) \eta^2\Sigma_{\rm s}/ \tau $.
Eliminating $\tau$ of fragments at the low-mass end of the cascade by
Eq.~(\ref{eq:tstop_est}), we get
\begin{equation}
 \frac{d\Sigma_{\rm s}}{dt} = - \left(\frac{9}{4}-q\right)\frac{\Sigma_{\rm s}h_M^3
  \langle P_{\rm VS} \rangle \Omega_{\rm K} \eta v_{\rm K}^2}{2^{1/3}
  C_{\rm L}\pi {\tilde b} Q_{\rm D}^*}.\label{eq:dsigma_gas} 
\end{equation}
Since $\langle P_{\rm col} \rangle \simeq 11.3 \sqrt{\tilde R}$ is
independent of $M$ for $e^* \ll h_M$, integration of Eq.~(\ref{eq:dm})
divided by Eq.~(\ref{eq:dsigma_gas}) gives another formula for the final
embryo mass, 
\begin{equation}
 M_{\rm f} = \left[
\frac{2^{7/3} C_{\rm L} {\tilde b} \pi C_{\rm acc} a^2 \langle P_{\rm col} \rangle
(3 M_*)^{1/3}}{ 3 (9/4-q) \eta P_{\rm VS}}
  \frac{Q_{\rm D}^*}{v_{\rm K}^2}(\Sigma_{\rm s,0}-\Sigma_{\rm s})
\right]^{3/4},\label{eq:mpla_frag}  
\end{equation}
where we set $M=0$ at $\Sigma_{\rm s} = \Sigma_{\rm s,0}$.  We recall
that $\langle P_{\rm VS} \rangle = 73$ and $\langle P_{\rm col} \rangle
= 11.3 \sqrt{\tilde R}$ in Eq.~(\ref{eq:mpla_frag}) are independent of
$M_{\rm f}$ for $e^* \ll h_M$.  In addition, since $Q_{\rm D}^*$ for the
strength regime is almost independent of the fragment mass $m$, we
neglected the mass dependence of $Q_{\rm D}^*$ for the derivation of
Eq.~(\ref{eq:mpla_frag}).  Assuming $\Sigma_{\rm s,0} \gg \Sigma_{\rm
s}$ and $q=3/2$, we obtain
\begin{eqnarray}
 M_{\rm f} &=& 0.14 
  \Biggl(\frac{\tilde b}{10}\Biggr)^{3/4}
  \Biggl(\frac{a}{5\,{\rm AU}}\Biggr)^{3/2}
  \Biggl(\frac{M_*}{M_\sun}\Biggr)^{3/8}
  \nonumber
  \\
 &&\times
  \Biggl(\frac{\rho}{1\,{\rm g/cm}^3}\Biggr)^{-1/8}
  \Biggl(\frac{\Sigma_{\rm s,0}}{2.7\,{\rm g/cm}^2}\Biggr)^{3/4}
  \Biggl(\frac{Q_{\rm D}^*}{3.1\times 10^6 \, {\rm erg/g}}\Biggr)^{3/4}
  M_{\oplus}.\label{eq:M_f_est} 
\end{eqnarray}
The final mass given by Eq.~(\ref{eq:M_f_est}) is also much smaller than
the isolation mass, Eq.~(\ref{eq:M_iso}).  Furthermore, $M_{\rm f}$ is
comparable to $M_{\rm c}$ given by Eq.~(\ref{eq:M_c_est}).  When
planetesimals have a large $Q_{\rm D}^*$ because of a rigid material
(high $Q_{\rm 0g}$) and/or a large size, $M_{\rm c}$ is larger than
$M_{\rm f}$.  In the opposite case, $M_{\rm f}$ exceeds $M_{\rm c}$.
The final mass is supposed to be the larger of the two, $M_{\rm c}$ and
$M_{\rm f}$.  We apply $Q_{\rm D}^*$ for $r=10$\,m and will compare
$M_{\rm f}$ with the embryo mass resulting from the statistical
simulation.

\section{NUMERICAL SIMULATIONS}
\label{eq:num}

\subsection{METHOD}

Many authors attacked the problem of the planetary growth from
planetesimals to embryos by $N$-body simulations (e.g., Kokubo and Ida
1996; 2002), the statistical method (Wetherill and Stewart 1993; Kenyon
and Bromley 2004), and the hybrid method (Kenyon and Bromley 2008,
Chambers 2008).  We apply the statistical method developed by Inaba et
al. (1999, 2001), which we briefly explain now.  The mass distribution
of bodies in orbit around a central star evolves through mutual
collisions, taking into account gravitational focusing of colliding
bodies (see Inaba et al. 2001 for detailed expressions).  Along with the
mass evolution, velocity dispersion changes by gravitational
perturbations, collisional damping, and gas drag.  Equations for the
mass distribution of bodies are integrated simultaneously with equations
for their velocity distribution.

The time evolution of the differential surface number density $n_{\rm s}
(m)$ at a distance $a$ is given by
\begin{eqnarray}
 \frac{\partial m n_{\rm s}(m,a)}{\partial t} &=& 
\frac{m}{2} \Omega_{\rm K} \int_0^m d m_1 
\int_{m-m_1-m_{{\rm e}}}^\infty d m_2 
\nonumber
\\ 
&& \quad \quad \times (h_{m_1,m_2} a)^2
n_{\rm s}(m_1,a) n_{\rm s}(m_2,a) \langle P_{\rm col} \rangle
\nonumber \\
&& \quad \quad \times \delta(m-m_1-m_2+m_{{\rm e}})
\nonumber
\\
&& - \Omega_{\rm K} m n_{\rm s}(m) \int_0^\infty d m_2 (h_{m,m_2} a)^2
n_{\rm s}(m_2,a) \langle P_{\rm col} \rangle
\nonumber
\\
&& +\frac{\partial}{\partial m} \Omega_{\rm K} \int_m^\infty d m_1\int_0^{m_1} d m_2
 (m_1+m_2) f(m,m_1,m_2) 
\nonumber
\\
 && \quad \quad \times n_{\rm s}(m_1,a) n_{\rm s}(m_2,a) (h_{m_1,m_2} a)^2
  \langle P_{\rm col}
  \rangle
\nonumber
\\
&& - \frac{1}{a} \frac{\partial}{\partial a} [a m n_{\rm s}(m,a) v_{\rm drift}(m,a)], 
\label{eq:mass_ev}
\end{eqnarray}
where $\delta(x)$ is the delta function, $(m_1+m_2) f(m,m_1,m_2)$ is the
mass of fragments less than $m$ produced by a collision between $m_1$
and $m_2$, and $m_{\rm e}$ is the total mass of the fragments (given by
Eq.~(\ref{eq:fragmass}) in our fragmentation model).  We apply the
dimensionless collisional rate $\langle P_{\rm col} \rangle$ which Inaba
et al.~(2001) provide as a function of $m_1$, $m_2$, and $e^*$ and $i^*$
by compiling previous studies.  Equation~(\ref{eq:mass_ev}) describes
the mass transport in the two-dimensional space composed of mass and
distance (radial direction).  The first and second terms in the
right-hand side of Eq.~(\ref{eq:mass_ev}) represent the mass transport
along the mass coordinate caused by coagulation and the third term does
that due to fragmentation. The fourth term describes the mass transport
along the radial coordinate due to the drift of bodies.  We calculate
the transport on a grid of mass and radial bins.

Assuming a power-law mass distribution of fragments, $(m_1+m_2)
f(m,m_1,m_2)$ is given by
\begin{equation}
 (m_1+m_2) f(m,m_1,m_2) = \left\{
  \begin{array}{lll}
\displaystyle 
   m_{{\rm e}}\left(\frac{m}{m_{{\rm L}}}\right)^{-b} 
    & {\rm for} & m < m_{{\rm L}},    
   \\
\displaystyle 
   m_{{\rm e}}
    & {\rm for} & m \geq m_{{\rm L}}, 
  \end{array}
  \right.\label{eq:functionf}
\end{equation}
where we recall the mass of 
the largest ejecta $m_{{\rm L}} = \epsilon (m_1+m_2) \phi /(1+\phi)^2$  
 and the total ejecta mass $m_{\rm e}$ given by Eq.~(\ref{eq:fragmass}). 

The drift velocity is characterized by the dimensionless parameter
$\tilde \tau_{\rm stop} $, where the scaled stopping time $\tilde
\tau_{\rm stop}$ is given by $a \Omega_{\rm K}^2 \tau / u$.  The
relative velocity $u$ between a body and gas is equal to $(e^* + i^* +
\eta) a \Omega_{\rm K}$.  For $\tilde \tau_{\rm stop} \la 1$ and $\tilde
\tau_{\rm stop} \gg 1$, the drift velocity can be written as (Adachi et
al. 1976, Inaba et al. 2001) 
\begin{equation}
 v_{\rm drift}(m,a) 
= 
\left\{
\begin{array}{l l l}
\displaystyle
 - \frac{2 \eta a \Omega_{\rm K}}{\tilde \tau_{\rm stop}} 
 \frac{\tilde \tau_{\rm stop}^2}{1+ \tilde \tau_{\rm stop}^2}
&{\rm for} &
\tilde \tau_{\rm stop} \la 1, 
\label{eq:vr_couple}
\\
 \displaystyle
  - 2 \frac{a \eta}{\tau}
  \left\{ \frac{[2E(3/4)+K(3/4)]^2}{9\pi^2} e^{*2} +\frac{4}{\pi^2} i^{*2}
   + \eta^2\right\}^{1/2 }
&{\rm for} &\tilde \tau_{\rm stop} \gg 1, 
\label{eq:vr_quad} 
\end{array}
\right.
\end{equation}
where $E$ and $K$ are the elliptic integrals.  Both regimes in
Eq.~(\ref{eq:vr_couple}) can be combined into
\begin{equation}
 v_{\rm drift}(m,a) = - 2 \frac{a \eta}{\tau} \frac{\tilde
  \tau_{\rm stop}^2}{1+ \tilde \tau_{\rm stop}^2} 
  \left\{ \frac{[2E(3/4)+K(3/4)]^2}{9\pi^2} e^{*2} +\frac{4}{\pi^2} i^{*2}
   + \eta^2\right\}^{1/2}.\label{eq:vr_mix}
\end{equation}
Indeed, if $\tilde \tau_{\rm stop} \la 1$, then $e^*$ and $i^*$ are
damped by the strong gas drag to $e^* \ll \eta$ and $i^* \ll \eta$, and
Eq.~(\ref{eq:vr_mix}) reduces to the first of Eqs.~(\ref{eq:vr_quad}).
If $\tilde \tau_{\rm stop} \gg 1$, Eq.~(\ref{eq:vr_mix}) simply
coincides with the second of Eqs.~(\ref{eq:vr_quad}).

Since we treat small fragments, we consider three gas-drag regimes:
quadratic, Stokes, and Epstein ones. The gas drag coefficients $C_{\rm
D}$ for these regimes are given by (Adachi et al. 1976)
\begin{eqnarray}
 C_{\rm D, quad} = 0.5, \quad C_{\rm D, Stokes} = 24/Re, \quad C_{\rm
  D, Epstein} = 16 /K Re, 
\end{eqnarray}
where $K = 1.7 \, (10^{-9} {\rm \, g \,cm}^{-2}/\rho_{\rm gas} r) $ is
the Knudsen number and $Re = 4.4 (u/c) K^{-1}$ is the Reynolds number
with $c$ being the sound velocity of gas (Adachi et al. 1976).  All
three regimes can be described together by adopting
\begin{equation}
 C_{\rm D} = C_{{\rm D, quad}} + \left(\frac{1}{C_{{\rm D, Stokes}}} +
				  \frac{1}{C_{{\rm D, Epstein}}} \right)^{-1}.\label{eq:cd_adopt} 
\end{equation}

Since the mean collision rate is a function of $e^*$ and $i^*$, the
evolution of $e^*$ and $i^*$ should be calculated precisely.  We
consider gravitational perturbations from other bodies, collisional
damping, and gas drag. Their net effect can be calculated as a square of
dispersions:
\begin{eqnarray}
\frac{d e^{*2}}{d t} &=& \left(\frac{de^{*2}}{dt}\right)_{\rm grav}
 + \left(\frac{de^{*2}}{dt}\right)_{\rm gas}
 + \left(\frac{de^{*2}}{dt}\right)_{\rm coll},\label{eq:de} \\
\frac{d i^{*2}}{d t} &=& \left(\frac{di^{*2}}{dt}\right)_{\rm grav}
 + \left(\frac{di^{*2}}{dt}\right)_{\rm gas}
 + \left(\frac{di^{*2}}{dt}\right)_{\rm coll},\label{eq:di} 
\end{eqnarray}
where terms with subscripts ``grav'', ``gas'', and ``coll'' indicate the
time variation due to gravitational perturbations, gas drag, and
collisions, respectively. These terms are provided by Inaba et
al. (2001).  Note that for the collisional damping, both fragments and
mergers resulting from a collision have the velocity dispersion at the
gravity center of colliding bodies, as described in their Eq.~(30).  We
follow the planetary growth by simultaneously integrating 
Eqs.~(\ref{eq:mass_ev}), (\ref{eq:de}), and (\ref{eq:di}).

Kokubo and Ida (1998) and Weidenschilling et al. (1997) showed that the
radial separation of orbits of runaway bodies formed in a swarm of
planetesimals is about 10 times their mutual Hill radius.  Since
dynamical friction from the field planetesimals makes orbits of the
runaway bodies nearly circular and coplanar, the orbital crossing of
these runaway bodies never occurs before the gas is dispersed (Iwasaki
et al. 2001).  Therefore, they cannot collide with each other.  In our
simulation, when the bodies reach a certain mass $m_{\rm run}$ such that
the sum of their mutual Hill radii equals the radial-bin width divided
by $\tilde b$, the bodies are regarded as ``runaway bodies'' which do
not have any collisions and dynamical interaction due to close
encounters among them, following Inaba et al. (2001).  We set $\tilde b
= 10$.  As the bodies grow, their separation measured in the Hill radii
decreases and these bodies can no longer be treated as ``runaway
bodies''.  Therefore, $m_{\rm run}$ increases during the simulation and
then the number of the ``runaway bodies'' decreases. 

Moreover, Inaba et al. (2001) set the number of collisions during a
numerical time interval to be an integer using a random number generator
and hence keep the number of bodies an integer.  Our procedure is
different.  Instead of dealing with the number of bodies, we treat the
mathematical expectation $N$ of the number of bodies with mass larger
than $m$.  When $N$ is much lower than unity, the bodies may not yet
exist.  We introduce a certain critical number $N_{\rm c}$ (a
``threshold'') to get rid of such bodies.  If $N(m_{\rm c}) = N_{\rm
c}$, we neglect collisions with bodies larger than $m_{\rm c}$ and
dynamical interactions with them.  The value of $N_{\rm c}$ should be of
the order of unity, because we can say that bodies are not yet born for
$N \ll 1$, whereas choosing a high $N_{\rm c}$ would delay the embryo
growth.  Bodies with masses ranging from $m_{\rm run}$ to $m_{\rm c}$
are treated as ``runaway bodies''.  The mass range of the ``runaway
bodies'' tends to extend during the simulation for low $N_{\rm c}$,
although being kept small in $N$-body simulation (Kokubo and Ida 2002).
We always start the simulation with $N_{\rm c} =0.1$ and $N_{\rm c}$ may
change from 0.1 to 10 during the simulation to keep the small mass range
of the ``runaway bodies''.  As shown in Figs.~\ref{fig:comp_mass_evo}
and \ref{fig:comp_ei_evo}, our simulation with such a choice of $N_{\rm
c}$ reproduces the mass and velocity-dispersion distributions obtained
from the $N$-body simulation, presented in Figs.~4 and 5 of Inaba et
al.~(2001).  Where about 100 runs of the statistical code by Inaba et
al.~(2001) are required to reproduce the $N$-body simulation, our
simulation does the same with only one run.

\subsection{EMBRYO GROWTH}

\subsubsection{WITHOUT FRAGMENTATION}

We calculate the evolution of the number and velocity dispersion of
bodies by summing up the time variations coming from all mass and radial
bins and by integrating them over time.  We integrate 
Eq.~(\ref{eq:mass_ev}) through the fourth order of Runge-Kutta method
for the mass evolution and Eqs.~(\ref{eq:de}) and (\ref{eq:di}) through
the linear method for $e^*$ and $i^*$ evolution.  We set the mass radio
between adjacent mass bins to 1.2, which we found sufficient to
reproduce $N$-body simulations (see Figs.~\ref{fig:comp_mass_evo} and
\ref{fig:comp_ei_evo}).  Fig.~\ref{fig:cumnum_nofrag} shows the results
ignoring fragmentation ($Q_{\rm D}^* = \infty$) with a set of six
concentric annuli at 3.2, 4.5, 6.4, 9.0, 13, and 18\,AU for $M_* =
M_\sun$, $\Sigma_1 = 7 \,{\rm g/cm}^2$, and $q=3/2$.  We consider
planetesimals with initial mass of $m_0 = 4.2\times 10^{18}\, {\rm g}$
(radius of $r_0 = 10\,$km for $\rho = 1\,{\rm g/cm}^3$) and velocity
dispersion given by $e^* = 2 i^* = (2 m_0/M_*)^{1/3}$.  The
planetesimals are assumed to be composed of ice whose physical
parameters are listed in Table 1.  We artificially apply the gas surface
density evolution in the form $\Sigma_{\rm gas} = \Sigma_{\rm
gas,0}\exp(-t/T_{\rm gas,dep})$, where $T_{\rm gas,dep}$ is the gas
depletion timescale\footnote{Assuming a constant $\Sigma_{\rm gas}$
gives almost the same results for the final embryo mass, because we
consider time spans $t \leq T_{\rm gas,dep}$.}.  Here, we set $T_{\rm
gas,dep} = 10^7$ years.

Figure~\ref{fig:cumnum_nofrag}a shows the mass distribution of bodies at
3.2\,AU from $10^4$ years to $4\times 10^6$ years.  After $10^5$ years,
the mass distribution of small bodies ($m \la 10^{24}$\,g) is a single
power law, which is consistent with that of the runaway growth, $n_{\rm
c} \propto m^{-5/3}$ (Makino et al. 1998).  At the same time, the
oligarchic growth results in the flatter distribution for large bodies
($m \ga 10^{24}$\,g).  Such large bodies have almost a single mass
($\sim 3 \times 10^{26}$\,g at $10^6$ years and $\sim 2 \times
10^{27}\,$g at $4\times 10^6$ years).  As the growth of the large bodies
proceeds, the random velocities of small bodies exceed the escape
velocity because of the viscous stirring by large bodies in the entire
mass range of small bodies ($m \la 10^{24}$\,g).  Then, the mass
distribution starts to depart from this power law.  Collisions between
small bodies at such velocities lead to a flat mass distribution for
small bodies ($m \la 10^{{21}}\,$g at $10^6$ years and $m\la 10^{22}$\,g
at $4 \times 10^6$ years) and a steep one for large bodies.  This means
that planetesimals dominating the surface density become larger after
$10^5$ years.

The velocity dispersion evolution ($e^*$ and $i^*$) at 3.2\,AU is shown
in Fig.~\ref{fig:vel_nofrag}. At $10^4$ years, $e^*$ and $i^*$ are
proportional to $m^{-1/2}$, being determined by the dynamical friction.
At $10^5$ years, $e^*$ and $i^*$ for $m \la 10^{22}$\,g are independent
of mass, whereas $e^*$ and $i^*$ of larger bodies still preserve an
$m^{-1/2}$ dependence.  The viscous stirring by large bodies dominates
$e^*$ and $i^*$ of small bodies. In this case, $e^*$ is given by
Eq.~(\ref{eq:e_nogas}).  After $10^6$ years, the velocity dispersion of
small bodies ($m \la 10^{22}\,$g at $10^6$ years and $m \la 10^{25}\,$g
at $4 \times 10^6$ years) reaches the equilibrium between the stirring
and the gas drag.  In this case, $e^*$ is given by Eq.~(\ref{eq:e_gas}).
However, bodies with intermediate mass do not yet have the velocity
dispersion in the equilibrium.  The velocity dispersion of bodies
dominating the surface density is still given by Eq.~(\ref{eq:e_nogas}).
The growth of large bodies are described by $M_{\rm n}$ given by
Eq.~(\ref{eq:m_nogas}) in $10^5$--$10^7$ years (see Appendix A).

In the outer region, the mass distribution and velocity dispersion
evolve in a similar way but in a longer timescale.  At $4\times 10^6$
years, the bodies with $m \ga 10^{24}$\,g grow through the oligarchic
mode at $a \la 9$\,AU.  Beyond 9\,AU, the largest bodies have not yet
reached $10^{24}$\,g by that time.

\subsubsection{WITH FRAGMENTATION}

We now take into account fragmentation, using $Q_{\rm D}^*$ in
Eq.~(\ref{eq:qd}).  Kenyon and Bromley (2009) suggested that the small
bodies coupled with gas ($\la 1\,$m) affect the embryo growth.  To treat
the coupled bodies, we set the smallest mass to be $4.2$\,g (radius of
1cm for $\rho = 1\,{\rm g/cm}^3$).

The mass distribution and velocity evolution that we calculated with
fragmentation are shown in Figs.~\ref{fig:mass_ev} and \ref{fig:vel_ei}.
During the first $10^5$ years, bodies grow through mutual collisions.
Fragments ($m \la m_0$) are not yet numerous because of the low velocity
dispersion (Fig. \ref{fig:vel_ei}). At $t = 10^6$ years, the runaway
growth occurs inside 6.4\,AU.  Similar to the case without
fragmentation, the cumulative number $n_{\rm c}$ is almost proportional
to $m^{-5/3}$ for $m \ga 10^{20} \sim 100 \,m_0$ and the distribution of
large bodies with $m \ga 10^{24}\,$g is flat.  Since the velocity
dispersion of planetesimals $\sim e^* v_{\rm K}$ exceeds their escape
velocity for $m \la 10^{24}\,$g, planetesimals dominating the surface
density become larger after $10^6$ years.  In contrast to the case
without fragmentation, the mass of bodies dominating the surface density
ceases at about $100\,m_0$, owing to a high collisional velocity $e^{*}
v_{\rm K} \ga \sqrt{Q_{\rm D}^*}$.  The slope of $n_{\rm c}$ for $m \sim
10^{11}$--$10^{20}\,$g ($r=0.1$--$10$\,km) is nearly $- (5+3p)/(6+3p)$,
which is typical of a collision cascade for $v^2/Q_{\rm D}^* \propto
m^{-p}$ (Kobayashi and Tanaka 2010).  The downward mass flux along the
mass coordinate is constant with mass in the collision cascade.  For $m
\la 10^{11}$\,g, the velocity dispersions ($e^*$ and $i^*$) are
effectively damped by the gas drag in the Stokes regime (See
Fig.~\ref{fig:vel_ei}).  Because the collisional energy ($\sim e^{*2}
v_{\rm K}^2$) is lower than the energy threshold ($Q_{\rm D}^*$) for a
destructive collision, the downward mass flux becomes insignificant
compared to that in the collision cascade.  Therefore, the number of the
bodies increases around $10^{11}$\,g where $e^{*2} v_{\rm K}^2 \sim
Q_{\rm D}^*$.  Bodies with $m\sim 10^5\,$g ($r\sim 1$\,m) are most
efficiently removed by the radial drift and smaller bodies ($m \ll
10^5$\,g or $r \ll 1$\,m) are coupled with gas. As a result, the mass
distribution becomes wavy (see Fig.~\ref{fig:mass_ev}). The surface
density of bodies with $e^{*2} v_{\rm K}^2 \sim Q_{\rm D}^*$ is much
higher than that of the coupled bodies.  However, planetesimals with
$\sim 100\,m_0$ dominate the surface density.  At $4\times 10^6$ years,
the oligarchic growth is inhibited by fragmentation inside 5\,AU, where
the number of fragments decreases compared to that at $10^6$ years.

Figure \ref{fig:emb_growth} shows the embryo-mass\footnote{Here, embryo
mass represents the average mass of the ``runaway bodies''.} evolution
at 3.2\,AU.  At the runaway growth phase (the first $\la 10^5$ years),
the embryo mass grows as an exponential function of time.  The growth
rate becomes much lower in the oligarchic mode at $10^5$--$10^6$ years.
The embryos exceed that for the case without fragmentation.  They are
substantially fed by small fragments whose velocities are damped by gas
drag and are low (see Figs.~\ref{fig:vel_ei}), resulting in their rapid
growth.  However, they grow at a sluggish pace at about $4 \times 10^5$
years because the surface density of planetesimals is decreased by
collision cascade followed by removal of small bodies by gas drag.  As a
result, the final embryo mass is reduced by fragmentation.  In this
case, since $M_{\rm c}$ and $M_{\rm f}$ are comparable, the embryo mass
is determined by both effects.  Even so, the final embryo mass is
consistent with $M_{\rm c}$ and $M_{\rm f}$.

Fig.~\ref{fig:final_mass} presents the embryo mass at $10^7$\,years as a
function of distance, including the results for three additional annuli
at 1, 1.4, and 2.0\,AU.\footnote{The results come from two simulations:
one in $a < 2.7\,$AU and the other in $a > 2.7$\,AU.}  In simulations at
$a \leq 2.7$\,AU, we use silicate bodies (Table~1) instead of icy
bodies.  Decline of the planetesimal surface density caused by
fragmentation results in embryos smaller than those for the case without
fragmentation inside 6\,AU.  The masses reach only $0.01$--0.1$M_\oplus$
in $10^7$ years.  The embryo masses at $a \la 1.5\,$AU and
$a=2.7$--$5$\,AU are consistent with $M_{\rm c}$ or $M_{\rm f}$.  Inside
$a \la 1.5$\,AU, the final mass is determined by the largest of the two
masses, $M_{\rm c}$ and $M_{\rm f}$ (see
Fig.~\ref{fig:emb_growth_at1au}).  For $m_0 = 4.2\times 10^{15}$\,g
($r_0 = 1\,$km), the embryo mass is determined by $M_{\rm f}$ because of
$M_{\rm f} > M_{\rm c}$ (see Fig.~\ref{fig:emb_growth_r0_1}).  In
addition, if both $M_{\rm c}$ and $M_{\rm f}$ are smaller than $M_{\rm
n}$ in Eq.~(\ref{eq:m_nogas}), the fragmentation is nearly negligible
and the embryo mass is given by $M_{\rm n}$ (see Appendix A for an
analysis of embryo growth without fragmentation).

Since the initial mass of planetesimals depends on their formation
process, which is not well understood, it is worthwhile investigating
the embryo mass dependence on the initial planetesimal
mass. Fig.~\ref{fig:fm_dep_r0} shows the dependence of the embryo masses
on $m_0$ at 3.2\,AU for the minimum mass solar nebula after 10 million
years.  If the initial mass is smaller than $\sim 10^{17}$\,g, the
embryo mass is determined by accretion of small fragments.  For larger
$m_0$, growth is halted by the collision cascade.  The embryo mass
increases with $m_0$ because the timescale of the surface density
decline due to collision cascade is longer for larger planetesimals,
strengthened by self-gravity (large $Q_{\rm D}^*$).  If $m_0 \ga
10^{21}$\,g, the embryo grows without substantial fragmentation.  For a
massive disk ($\Sigma_1 = 70\,{\rm g/cm}^2$), the embryo mass can reach
the Earth mass. In this case, $M_{\rm c}$ and $M_{\rm f}$ determine the
embryo mass because of the short growth time for $r_0 = 1$--100\,km.

\section{SUMMARY AND DISCUSSION}

In this paper, we studied analytically and numerically the growth of planetary
embryos in the framework of the standard scenario. We took into account 
that embryos growing in the oligarchic mode pump up relative velocities
of planetesimals, which causes their collisional fragmentation. We also
considered the fate of smallest fragments of the resulting collision cascade,
namely their inward drag in the ambient gas and possible accretion by nascent 
embryos.

Taking into account fragmentation, we have analytically derived the
final mass of a planetary embryo in Section \ref{sec:theo} in two cases:
the mass $M_{\rm c}$ of an embryo growing through the accretion of
planetesimals which are removed by collision cascade and the mass
$M_{\rm f}$ in the case where an embryo grows through collisions with
fragments which are removed by gas drag.  In Sec.~\ref{eq:num}, we
showed that the final embryo mass obtained in numerical simulations can
be reproduced by the larger of the two analytic estimates, MAX($M_{\rm
c}, M_{\rm f}$).  If the embryo mass $M_{\rm n}$ of
Eq.~(\ref{eq:m_nogas}) for the case without fragmentation is smaller
than MAX($M_{\rm c}, M_{\rm f}$), fragmentation is negligible and the
final mass is given by $M_{\rm n}$.  Altogether, our analytical formulae
provide a good estimate of the final mass of planetary embryos.

The gas giant planet formation via core accretion requires the solid
core as large as $\sim 10 M_{\oplus}$ (e.g., Ikoma et al.~2001).  Kenyon
and Bromley (2009) suggested that the planetary embryo may further grow
by the accretion of bodies coupled with gas because the coupled bodies
are no longer involved in collision cascade.  Taking into account three
regimes of gas drag (see Eq.~\ref{eq:cd_adopt}), we found that the
collision cascade halts at bodies having $v^2 \sim Q_{\rm D}^*$, i.e. at
sizes that are larger than the critical coupling size corresponding to
$\tilde{\tau}_{\rm stop} = 1$.  As a result, coupled bodies should be
produced only in little amounts and can hardly contribute to planetary
growth, despite their low velocity dispersion in the laminar disks
considered here.

In this paper, we derive the final embryo masses, assuming a uniform
distribution of solid bodies.  However, the planetary embryos can open
gaps in a solid disk, which would affect the accretion of fragments on
the embryo and their removal by the gas drag.  Tanaka and Ida (1999)
derived a condition for the gap opening around a migrating
embryo. Replacing the embryo's migration time by the drift time
$\tau/2\eta^2$ for fragments, we can obtain a criterion for an embryo
with mass $M$ to open a gap in the solid disk from their Eq.~(3.6),
\begin{equation}
 \frac{\tau \Omega_{\rm K} h_M^2}{4 \pi \eta^2} \geq 0.81 
  \left[\sqrt{1+0.45(\tau\Omega_{\rm K}/2\pi)^{2/3}} +1 \right]^2,\label{eq:gap_crit} 
\end{equation}
where $\tau$ is given by Eq.~(\ref{eq:tau_gas}) for fragments
surrounding the embryos.  Since Eq.~(\ref{eq:gap_crit}) requires
$\tau\Omega_{\rm K}/2\pi \ll 1$ for $M \ga 0.1 M_\oplus$, the critical
fragment radius $r_{\rm g}$ above which Eq.~(\ref{eq:gap_crit}) is
satisfied is given by
\begin{equation}
 r_{\rm g} \simeq 1.0 \times 10^2 \left(\frac{M}{0.1
			      M_{\oplus}}\right)^{-1/3}
 \Biggl(\frac{a}{5\,{\rm AU}}\Biggr)^{3/8}\, \rm{m}, \label{eq:rg} 
\end{equation}
where we adopt $q=3/2$ and $C_{\rm D} = C_{\rm C, Stokes}$.  Levison et
al.~(2010) performed $N$-body simulations involving fragments and found
$r_{\rm g} \sim 30$\,m for $M \geq M_{\oplus}$, which is consistent with
Eq.~(\ref{eq:rg}). On the other hand, as we have shown, the solid
surface density is reduced by the elimination of fragments at the
low-mass end of collision cascade due to gas drag.  In addition, the
embryo grows through the accretion of the fragments for small initial
planetesimals.  The radius $r_{\rm e}$ of the fragments is estimated by
Eq.~(\ref{eq:tstop_est}) as
\begin{equation}
 r_{\rm e} \simeq 5.0 \left(\frac{M}{0.1 M_{\oplus}}\right)^{-1/2} 
  \Biggl(\frac{a}{5\,{\rm AU}}\Biggr)^{1/2}\, \rm{m}.\label{eq:re} 
\end{equation}
Eqs.~(\ref{eq:rg}) and (\ref{eq:re}) give $r_{\rm e} \ll r_{\rm g}$.
This means that the collision cascade quickly grinds planetesimals down
to the size much less than $r_{\rm g}$. As a result, the gap formation
in the solid disk composed of fragments with the radius $r_{\rm e}$ does
not occur and the embryo growth would not be influenced.

Considering the range of initial planetesimal radii $r_0 = 1$--100\,km,
and the range of disk surface densities $\Sigma_{1} \sim 7$--$70\,{\rm
g/cm}^2$ at 3.2\,AU (1--$10\times$MMSN), our results suggest that the
fragmentation averts the planetary growth at $M \sim
0.1$--$10\,M_\oplus$ at several AU.  This is consistent with the results
of planetary growth simulations beyond 30\,AU by Kenyon and Bromley
(2008).  The cores approaching the critical mass of $10\, M_\oplus$ can
only form at $a=3$--4\,AU, only from planetesimals that have $r_0 \sim
100\,$km initially, and only in densest nebula with $\Sigma_{1} \sim
70\,{\rm g/cm}^2$.  Thus it appears problematic to explain formation of
giant planets at 5--10\,AU, such as Jupiter and Saturn in the solar
system, even for $\Sigma_{1} \sim 70\,{\rm g/cm}^2$ and $r_0 \sim
100$\,km.

An effect that could help further is the enhancement of the embryo's
collisional cross section due to a tenuous atmosphere acquired by the
embryo in the ambient gas.  Inaba and Ikoma (2003) and Inaba et
al. (2003) have shown that a core with mass exceeding about $0.1
M_\oplus$ could grow by that effect to the critical core mass.  As found
here, the embryo masses can indeed be about $0.1 M_\oplus$ under the
minimum mass solar nebula conditions, at least for an initial
planetesimal radius greater than $10$\,km.

On any account, our results strongly suggest that increasing the disk
surface density and/or the initial embryo size helps forming larger
cores.  This brings up the question whether the values that we need to
grow the embryo to the critical mass, $10\times$MMSN and $r_0 \sim
100$\,km~--- or less, if the atmospheric effect is taken into
account~--- appear plausible.  As far as the disk density is concerned,
the answer is probably yes.  $10\times$MMSN is close to the
gravitational stability limit, and all densities below this limit cannot
be ruled out. For instance, Desch (2007) and Crida (2009) point out that
the MMSN profile is inconsistent with the ``Nice model'' (Gomes et
al. 2004) and should be replaced with another surface density profile,
which would possibly imply larger surface densities at several AU.

We now consider the initial size of planetesimals. The mechanisms of
planetesimal formation are highly debated but, despite intensive effort,
remain fairly unknown. Classical models in which dust smoothly grows to
planetesimals with $r_0 \sim 1$\, km face the ``meter barrier''
problems: first, meter-sized objects should be lost to the central star
as a result of gas drag (Weidenschilling 1977, Brauer et al. 2008), and
second, further agglomeration of meter-sized objects upon collision is
problematic (Blum and Wurm 2008).  Accordingly, in the last years,
competing scenarios were suggested that circumvent these barriers, such
that the ``primary accretion'' mechanism proposed by Cuzzi et al. (2008)
and ``graviturbulent'' formation triggered by transient zones of high
pressure (Johansen et al. 2006) or by streaming instabilities (Johansen
et al. 2007).  These models all imply rapid formation of rather large
planetesimals in the $r_0 \sim 100$--$1000$\, km range.  Support for
these scenarios may come from the analysis of left-over planetesimals in
the solar system.  Morbidelli et al. (2009), for instance, suggest that
the initial planetesimals should be larger than 100\,km to reproduce the
mass distribution of asteroids in the main belt.

Our last remark is related to the Type I migration of bodies due to
interaction with gas (e.g., Tanaka et al. 2002).  The planetary embryo
grows in the runaway mode followed by the oligarchic one.  The timescale
of the runaway growth is proportional to the initial size of bodies. In
addition, the embryos rapidly grow through collisions with small
fragments for small $r_0$ even in the the oligarchic growth.  The embryo
eventually grows to a larger mass if it is large initially, while it
forms earlier if it starts with a smaller mass.  Therefore, initially
small planetesimals seem to be preferred for the core formation prior to
their removal due to Type I migration.  However, this is only valid for
laminar disks.  Turbulence may help prevent the loss of growing embryos
to Type I migration (e.g., Laughlin et al. 2004).  So, the alternative
scanarios of planetesimal formation that all require turbulent disks may
help here, too.

\section*{ACKNOWLEDGMENT}

We thank Glen R. Stewart and an anonymous referee for their thorough
reviews of this paper.  H. K. and A. V. K. are grateful to Martin Ilgner for helpful
comments.

\newpage 
\appendix

\section{PLANETARY GROWTH WITHOUT FRAGMENTATION}

When the growth timescale $T_{\rm grow}$ of embryos is much longer than
the $e^*$-damping timescale $T_{\rm gas}$ by gas drag, $e^*$ settles to
the equilibrium between the stirring and the gas drag and is given by
Eq.~(\ref{eq:e_gas}).  However, the growth timescale at the beginning of
oligarchic growth is relatively short.  In this case, $e^*$ cannot reach
the equilibrium for $T_{\rm grow} \ll T_{\rm gas}$.  Using these two
limits of $e^*$, we will derive the embryo mass evolution as follows.

In the non-equilibrium case ($T_{\rm grow} \ll T_{\rm gas}$), $e^*$
increases as $M$ grows.  Then, combining the growth rate
(Eqs.~(\ref{eq:dm})--(\ref{eq:pcol_high})) with the stirring rate
(Eq.~(\ref{eq:stiring}) with Eqs.~(\ref{eq:nM}) and
(\ref{eq:pvs_high})), we get
\begin{equation}
 e^* = 
\left[
 \frac{3 C_{\rm VS} \ln(\Lambda^2+1) M^{4/3}}{2^{10/3} \pi C_{\rm acc} \tilde b
 a^2 \Sigma_{\rm s} \tilde R C_{\rm col} (3 M_*)^{1/3}}
\right]^{1/2}, \label{eq:e_nogas} 
\end{equation}
where we assume that the initial embryo mass and eccentricity dispersion
are much smaller than $M$ and $e^*$, respectively.  Inserting
Eq.~(\ref{eq:e_nogas}) into Eq.~(\ref{eq:dm}) with
Eq.~(\ref{eq:pcol_high}) there results
\begin{equation}
 \frac{dM}{dt} = \frac{ 2^{10/3} \pi {\tilde b}
\left(C_{\rm acc} \Sigma_{\rm s} C_{\rm col} a^2 {\tilde R}\right)^2 \Omega_{\rm K} 
}{9 M_* C_{\rm VS} \ln (\Lambda^2 +1)}. 
\end{equation}
If the embryo grows with this rate for sufficiently long time, it is
valid use $M=0$ at $t=0$ as the initial condition.  In this case, the
embryo mass $M_{\rm n}$ at time $t$ is given by
\begin{equation}
 M_{\rm n} = \frac{ 2^{10/3} \pi {\tilde b}
\left(C_{\rm acc} \Sigma_{\rm s} C_{\rm col} a^2 {\tilde R}\right)^2 \Omega_{\rm K} t
}{9 M_* C_{\rm VS} \ln (\Lambda^2 +1)},\label{eq:m_nogas} 
\end{equation}
which is independent of $m$. 

In the equilibrium case ($T_{\rm grow} \gg T_{\rm gas}$), the growth
rate of $M$ is given by Eq.~(\ref{eq:dm_gas}).  Also we consider that
the embryo mass becomes much larger than the initial mass.  Integrating
Eq.~(\ref{eq:dm_gas}) over time with $M=0$ at $t=0$, we have
\begin{equation}
 M_{\rm e} = 
\Biggl[
\frac{C_{\rm acc} \Sigma a^2 C_{\rm col} {\tilde R} \Omega_{\rm
K}}{3 (3 M_*)^{2/3}}
\Biggr]^3
 \Biggl[
\frac{2^{4/3}\pi {\tilde b} C_{\rm gas}}{C_{\rm VS} \ln (\Lambda^2+1)
\Omega_{\rm K} \tau} 
  \Biggr]^{6/5} t^3.\label{eq:m_gas} 
\end{equation}
The embryo mass is proportional to $m^{2/5}$ from the $m$ dependence of
$\tau$.  The growth time of embryo mass $M_{\rm e}$ is similar to that
estimated by Kokubo and Ida (2002).

Fig.~\ref{fig:emb_growth_nofrag} shows the embryo growth at 3.2\,AU.
The runaway growth swithes to oligarchic growth after $10^5$ years.  The
embryo mass is consistent with $M_{\rm n}$ given by
Eq.~(\ref{eq:m_nogas}) in $10^5$--$10^7$ years.  On contrary, the embryo
mass does not agree with $M_{\rm e}$ given by Eq.~(\ref{eq:m_gas}). That
is because $e^*$ of the representative planetesimal, which dominates the
surface density, does not reach the equilibrium.  As seen in
Fig.~\ref{fig:cumnum_nofrag}, the representative planetesimal grows
because its random velocity is higher than its surface escape velocity.
Although $e^*$ reaches the equilibrium for small bodies, that of the
representative planetesimal cannot do due to the growth.  Since the
representative planetesimal growth is caused by the perfect sticking
even for a high velocity, the growth came from the artificial treatment.
In practice, the growth halts at $\sim 100\,m_0$ when fragmentation is
included, because the impact energy exceeds $Q_{\rm D}^*$.  In this
case, $e^*$ of the representative planetesimals reach the equilibrium
and the embryo mass is determined by $M_{\rm c}$ or $M_{\rm f}$ because
of the active fragmentation.

\section{MASS DEPLETION DUE TO COLLISION CASCADE}

Here, we derive the surface-density depletion rate due to collision
cascade, following Kobayashi and Tanaka (2010).  Focusing on the mass
transport by fragmentation, we neglect terms in the right-hand side of
Eq.~(\ref{eq:mass_ev}) except for the third one:
\begin{eqnarray}
 \frac{\partial m n_{\rm s}}{\partial t} &=& \frac{\partial}{\partial m}\Omega_{\rm K} \int_m^\infty d m_1\int_0^{m_1} d m_2
 (m_1+m_2) f(m,m_1,m_2) 
\nonumber
\\
 && \quad \times n_{\rm s}(m_1,a) n_{\rm s}(m_2,a) (h_{m_1,m_2} a)^2 \langle P_{\rm col}
  \rangle_{m_1,m_2}\label{eq:surfacedensity_evo1} 
\end{eqnarray}
Collisional velocities exceed the surface escape velocity of colliding
bodies in collision cascade, resulting in $(h_{m_1,m_2} a)^2 \langle
P_{\rm col} \rangle = (17.3/2 \pi) (r_1 + r_2)^2 = h_0
m_1^{2/3}[1+(m_2/m_1)^{1/3}]^2$.  Here, $r_1$ and $r_2$ are the radii of
$m_1$ and $m_2$, respectively.  Since collisions with $m_1 \gg m_2$
mainly lead to the mass transport along the mass coordinate in collision
cascade, the $f$ function given by Eq.~(\ref{eq:functionf}) becomes
\begin{equation}
 f(m,m_1,m_2) = \left\{
  \begin{array}{lll}
\displaystyle 
 \left(\frac{m}{m_1}\right)^{2-b} \frac{\phi}{1+\phi} \left[\frac{\epsilon \phi}{(1+\phi)^2}\right]^{b-2}
    & {\rm for} & \frac{m}{m_1} < \frac{\epsilon \phi}{(1+\phi)^2},
   \\
\displaystyle 
  \frac{\phi}{1+\phi}
    & {\rm for} & \frac{m}{m_1} \geq \frac{\epsilon \phi}{(1+\phi)^2}.
  \end{array}
  \right.
\end{equation}
Furthermore, $(h_{m_1,m_2} a)^2 \langle P_{\rm col} \rangle = h_0
m_1^{2/3}$ and $\phi = (v(m_1)^2/2 Q_{\rm D}^*(m_1)) (m_2/m_1)$.
Introducing the dimensionless variable $x = m/m_1$, we can re-write
Eq.~(\ref{eq:surfacedensity_evo1}) as
\begin{eqnarray}
 \frac{\partial m n_{\rm s}}{\partial t} &=& \frac{\partial}{\partial
  m}A^2 \Omega_{\rm K} h_0 
  m^{\frac{11}{3}-2\alpha} \left(\frac{v^2(m)}{2 Q_{\rm D}^*(m)}\right)^{\alpha-1}
\int_0^{v^2(m_1)/2Q_{\rm D}^*(m_1)} d \phi \frac{\phi^{1-\alpha}}{1+\phi}
\nonumber
\\
 && \quad \times \left[
\int_0^{\epsilon \phi / (1+\phi)^2} dx \, x^{2\alpha-8/3+(\alpha-1)p-b} 
\left(\frac{\epsilon \phi}{(1+\phi)^2}\right)^{b-2}
\right.
\nonumber
\\
&& \quad \quad \left.
+\int_{\epsilon \phi / (1+\phi)^2}^1 d x \, x^{2\alpha-14/3+(\alpha-1)p}
\right],\label{eq:cc_pri} 
\end{eqnarray}
where $n_{\rm s} = A m^{-\alpha}$ and $v^2(m)/Q_{\rm D}^*(m) \propto
m^{-p}$.  The upper integration limit over $\phi$ in
Eq.~(\ref{eq:cc_pri}) is different from Eq.~(31) of Kobayashi and Tanaka
(2010), which comes from the different definition of $f$.  Despite that,
Eq.~(\ref{eq:sigma_dep_collision_cascade}) is consistent with Kobayashi
and Tanaka because of a insignificant contribution of collisions with
$\phi > v^2 (m_1)/2 Q_{\rm D}^*(m_1)$ to the integral over $\phi$ in
their Eq.~(31).  Since $v^2(m_1)/2 Q_{\rm D}^*(m_1) \gg 1$, we change
the upper limit to $\infty$.  We also take into account that
$v^2(m)/Q_{\rm D}^*(m) \propto m^{-p}$ and that $\partial m n_{\rm
s}/\partial t = 0$ in a steady-state collision cascade.  Then,
Eq.~(\ref{eq:cc_pri}) gives $\alpha = (11+3p)/(6 +3p)$.  Note that
Eqs.(31), (34), (B2), and (B3) of Kobayashi and Tanaka (2010) contain
mistakes: their power-law exponents of $m$ should be $11/3 -2\alpha$, as
in Eq.~(\ref{eq:cc_pri}).

In the oligarchic growth, since planetary embryos sufficiently increase
the collisional velocity of the swarm planetesimals, collision cascade
depletes the swarm. Smaller bodies quickly reach the steady state of
collision cascade. Therefore, integrating Eq.~(\ref{eq:cc_pri}) over
mass in the steady-state collision cascade ($\alpha = (11+3p)/(6+3p)$),
we obtain
\begin{eqnarray}
 \frac{\partial \Sigma_{\rm s}}{\partial t} &=& - A^2 m_{\rm max}^{11/3-2\alpha} \Omega_{\rm K} h_0
  \left(\frac{v^2(m_{\rm max})}{2Q_{\rm D}^*(m_{\rm max})}\right)^{\alpha-1}
\int_0^\infty d \phi \frac{\phi^{1-\alpha}}{1+\phi}
\nonumber
\\
 && \quad \times \left[
\int_0^{\epsilon \phi/(1+\phi)^2} dx \, x^{1-b} 
\left(\frac{\epsilon \phi}{(1+\phi)^2}\right)^{b-2}
+
\int_{\epsilon \phi/(1+\phi)^2}^1 d x  \, x^{-1}
\right], \label{eq:deplete_ape}
\end{eqnarray}
where $m_{\rm max}$ is the mass of largest planetesimals in collision
cascade.  Integrating over $x$ on the right-hand side of
Eq.~(\ref{eq:deplete_ape}), and using $\Sigma_{\rm s} = A m_{\rm
max}^{2-\alpha} /(2-\alpha)$, we obtain
Eq.~(\ref{eq:sigma_dep_collision_cascade}).  In
Eq.~(\ref{eq:sigma_dep_collision_cascade}), we convert $m_{\rm max}$ to
$m$, according to the definition in Section \ref{sec:theo}.  Moreover,
Eq.~(\ref{eq:sigma_dep_collision_cascade}) includes the additional terms
due to the mass transport of the remnant according to Kobayashi and
Tanaka, although the transport terms are much smaller than
others.\footnote{The negative sign in the right-hand side of Eq.~(A5) of
Kobayashi and Tanaka (2010) is in error, and we correct this in
Eq.~(\ref{eq:sigma_dep_collision_cascade}).}

\newpage 

\newpage



\begin{center}
\renewcommand{\baselinestretch}{1}
\begin{table}[htbp]
 \caption{
Material properties
}
\begin{tabular}{c|c c c c c c}
 & $Q_{\rm 0s}$& ${\beta_{\rm s}}$ &
$Q_{\rm 0g}$ & ${\beta_{\rm g}}$ & $C_{\rm gg}$ & $\rho$\\
&(erg/g)&&(erg\,${\rm cm}^3/{\rm g}^2$)&&&(${\rm g/cm}^{3}$)\\
\hline 
 ice &  $7.0 \times 10^7$&  $-0.45$&  $2.1$& $1.19$& 9 & 1 \\
 silicate & $3.5\times 10^7$& $-0.38$& 0.3& $1.36$& 9 & 3
\end{tabular}
\end{table}
\renewcommand{\baselinestretch}{2}
\end{center}


\newpage

\begin{figure}[htbp]
\epsscale{0.7} \plotone{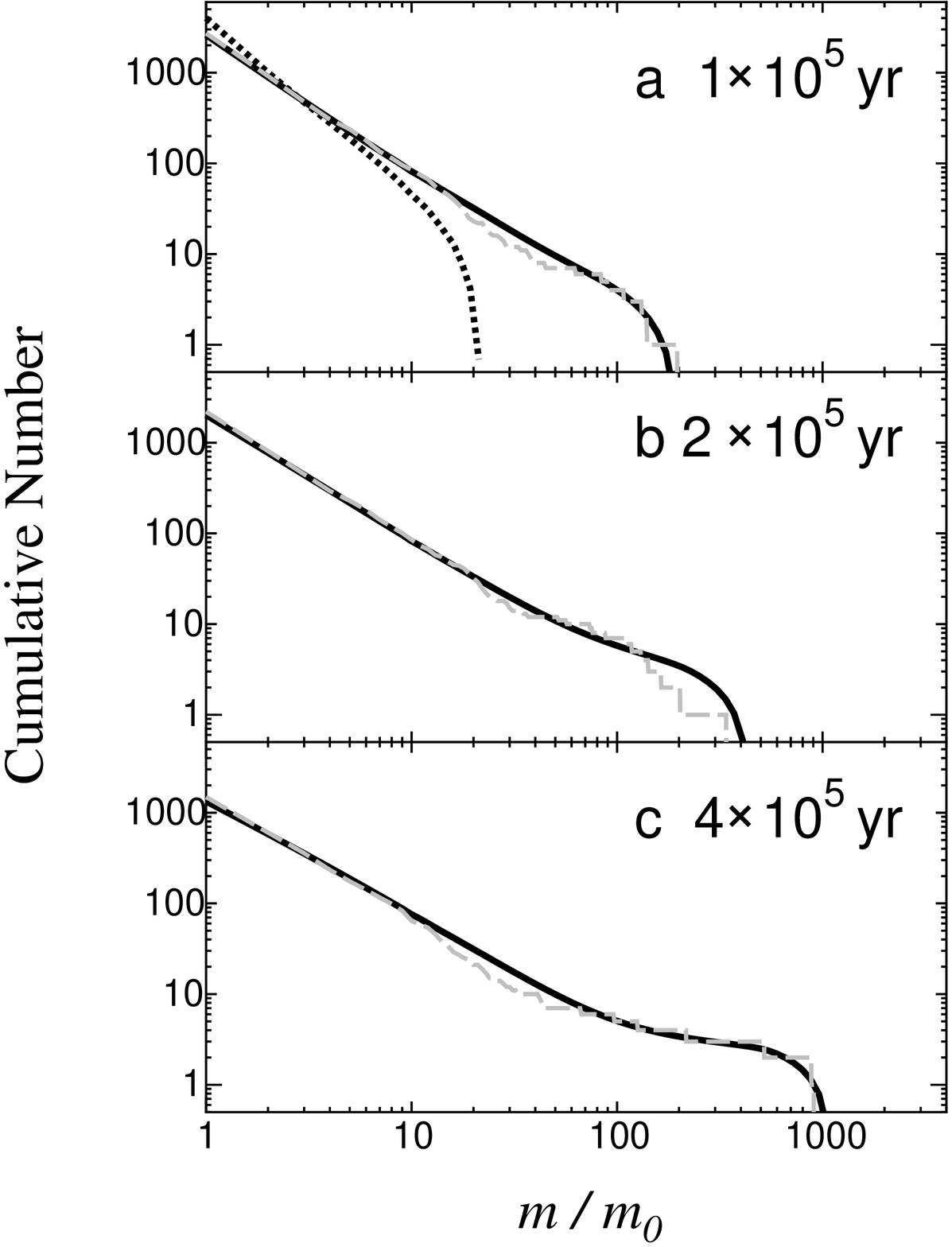} \figcaption{
\label{fig:comp_mass_evo}}
\end{figure}

Fig.\ \ref{fig:comp_mass_evo} --- Evolution of the mass distribution of
bodies with $m_0 = 2\times 10^{23}\,$g, $\rho = 2\,{\rm g/cm}^2$, and
$e^{*}_0 = 2 i^*_0 = 0.002$ at 1\,AU.  The initial mass distribution is
shown with the dotted curve in panel (a).  Collisional fragmentation is
neglected.  Gray dashed lines represent the results of $N$-body
simulation shown in Fig. 4 of Inaba et al. (2001).

\begin{figure}[htbp]
\epsscale{0.7} \plotone{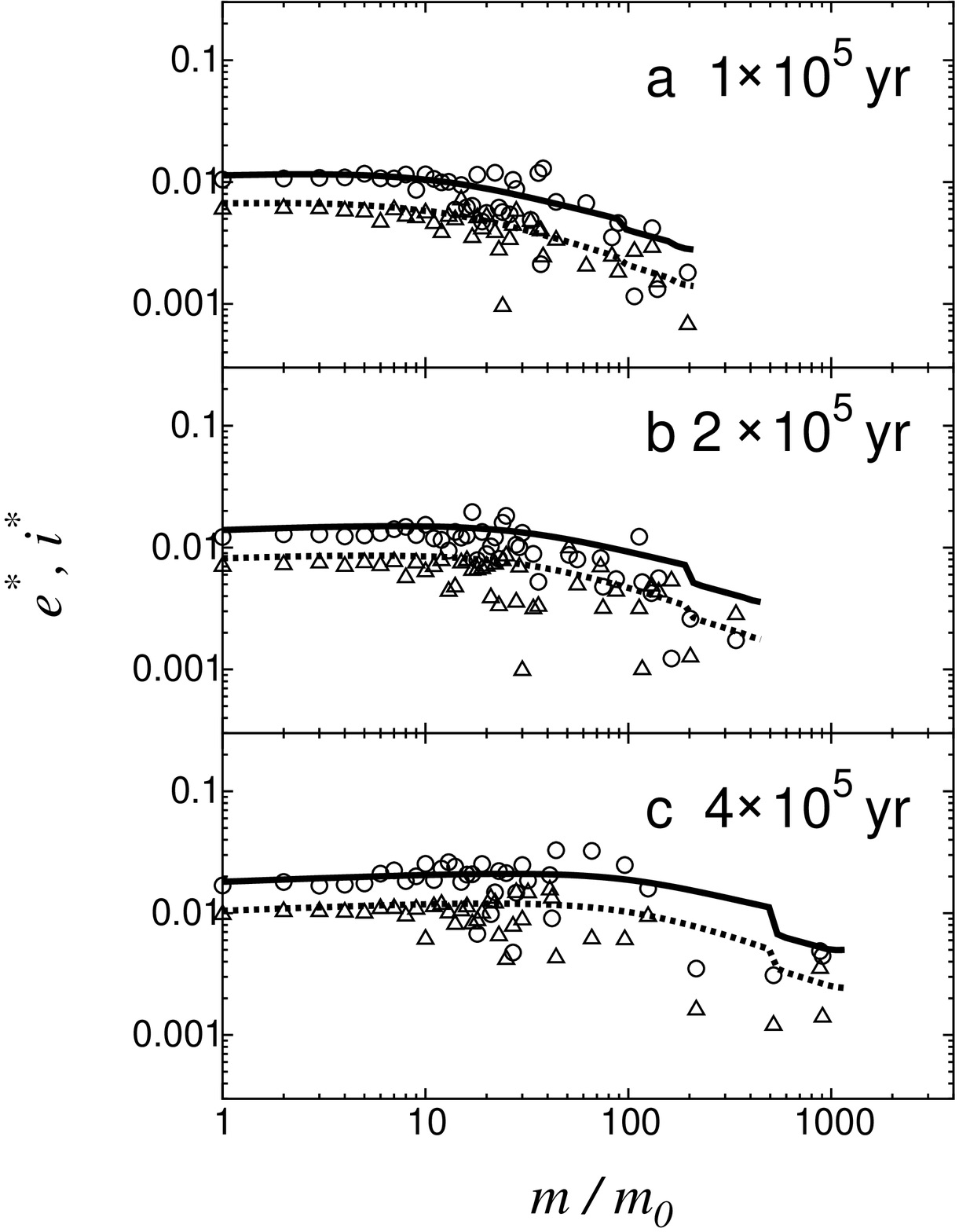} 
\figcaption{
\label{fig:comp_ei_evo}}
\end{figure}

Fig.\ \ref{fig:comp_ei_evo} --- Evolution of the velocity dispersion of
bodies for the same conditions as in Fig.~\ref{fig:comp_mass_evo}.  The
solid and dotted lines indicate $e^*$ and $i^*$, respectively.  Circles
and triangles represent $e^*$ and $i^*$ calculated by $N$-body
simulation shown in Fig. 4 of Inaba et al. (2001), respectively.

\begin{figure}[htbp]
\epsscale{0.9} \plotone{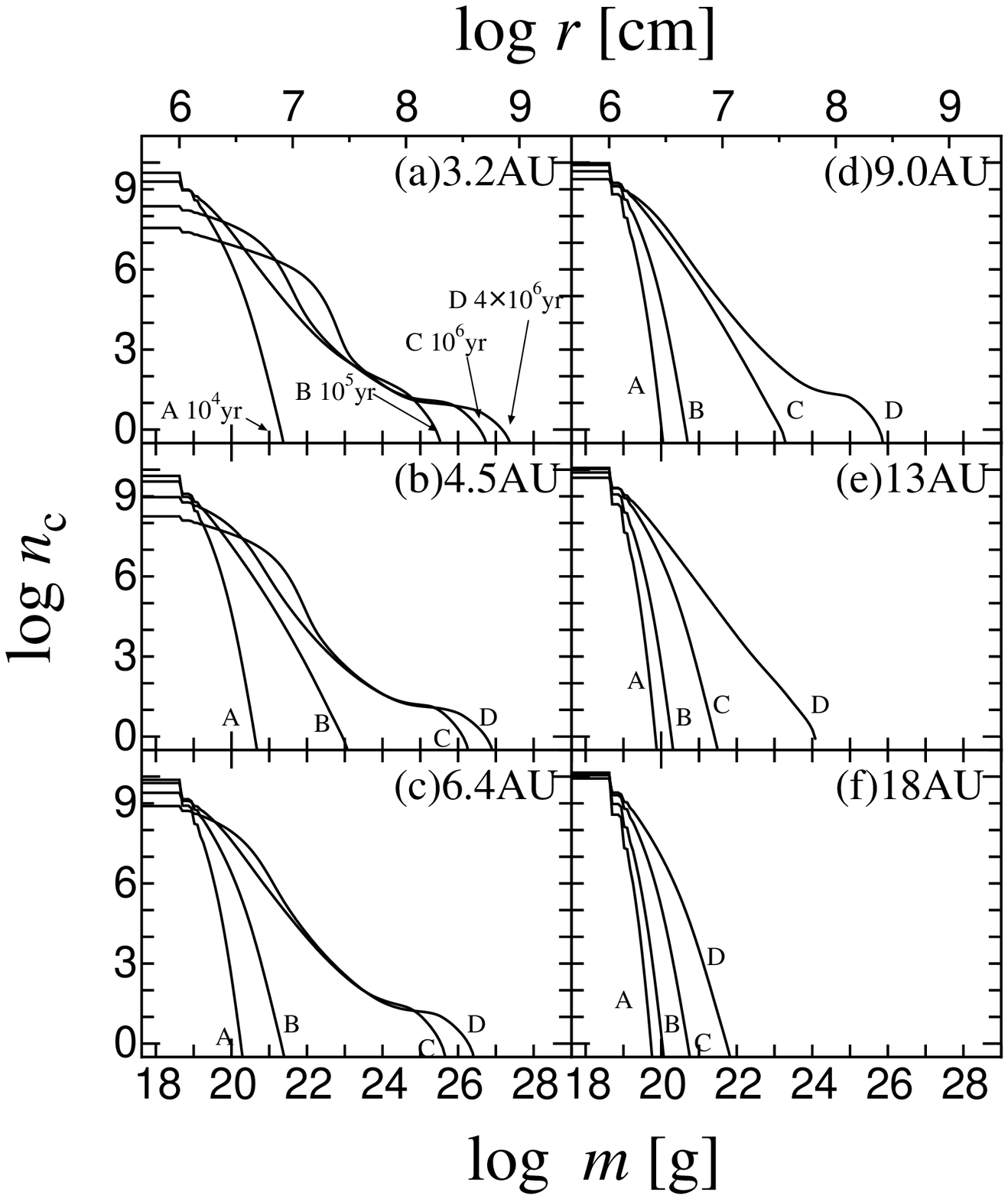} \figcaption{
\label{fig:cumnum_nofrag}}
\end{figure}

Fig.\ \ref{fig:cumnum_nofrag} --- The mass distribution of bodies at
$10^4$(A), $10^5$ (B), $10^6$ (C), $4\times 10^6$(D) years, neglecting
fragmentation.  Different panels correspond to different radial annuli.

\begin{figure}[htbp]
\epsscale{0.9} \plotone{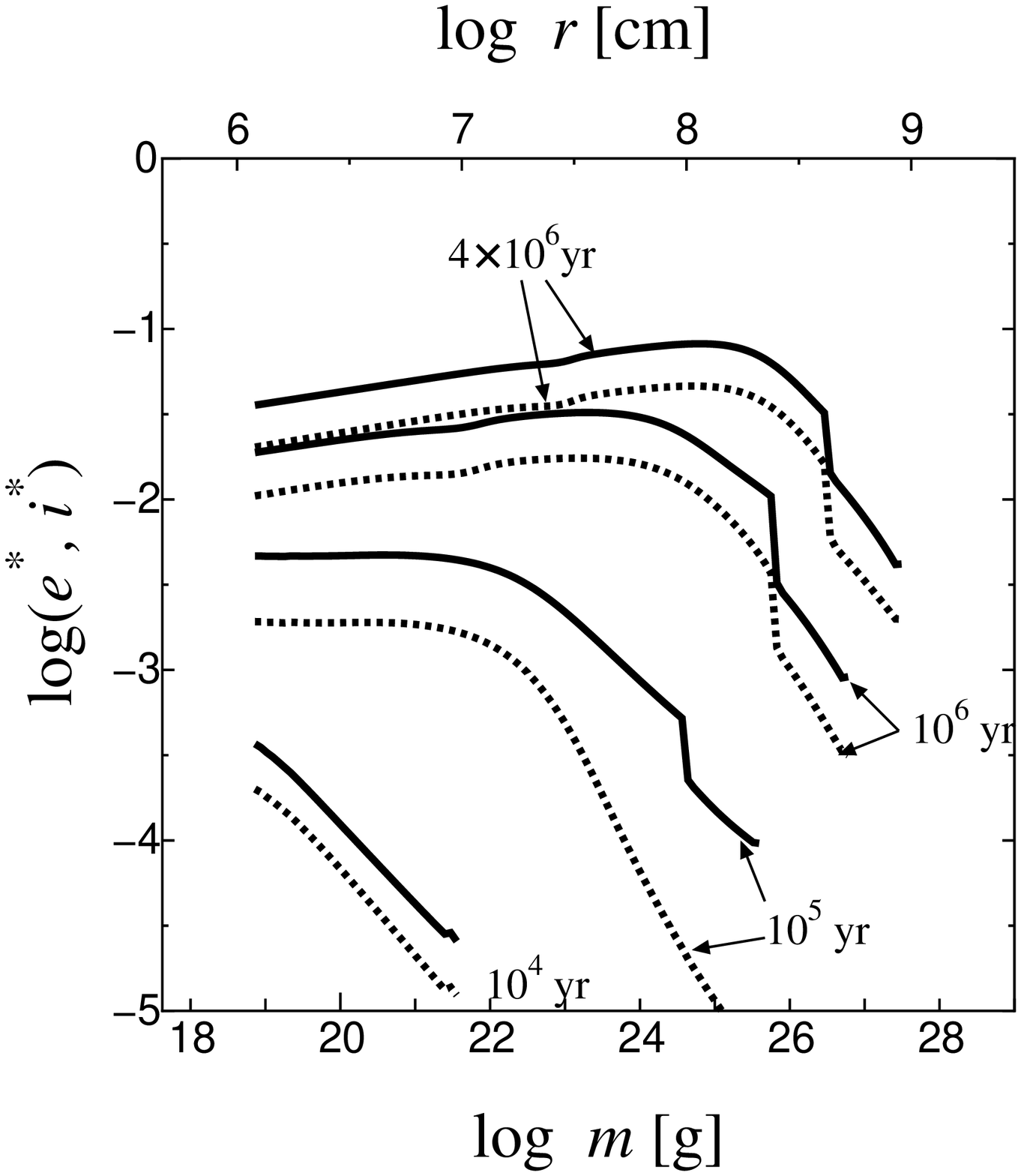} \figcaption{
\label{fig:vel_nofrag}}
\end{figure}

Fig.\ \ref{fig:vel_nofrag} --- The velocity-dispersion distribution at
3.2\,AU after $10^4$, $10^5$, $10^6$, $4\times 10^6$ years of evolution.
Fragmentation is neglected.  The solid and dotted lines indicate $e^*$
and $i^*$, respectively.

\begin{figure}[htbp]
\epsscale{0.9} \plotone{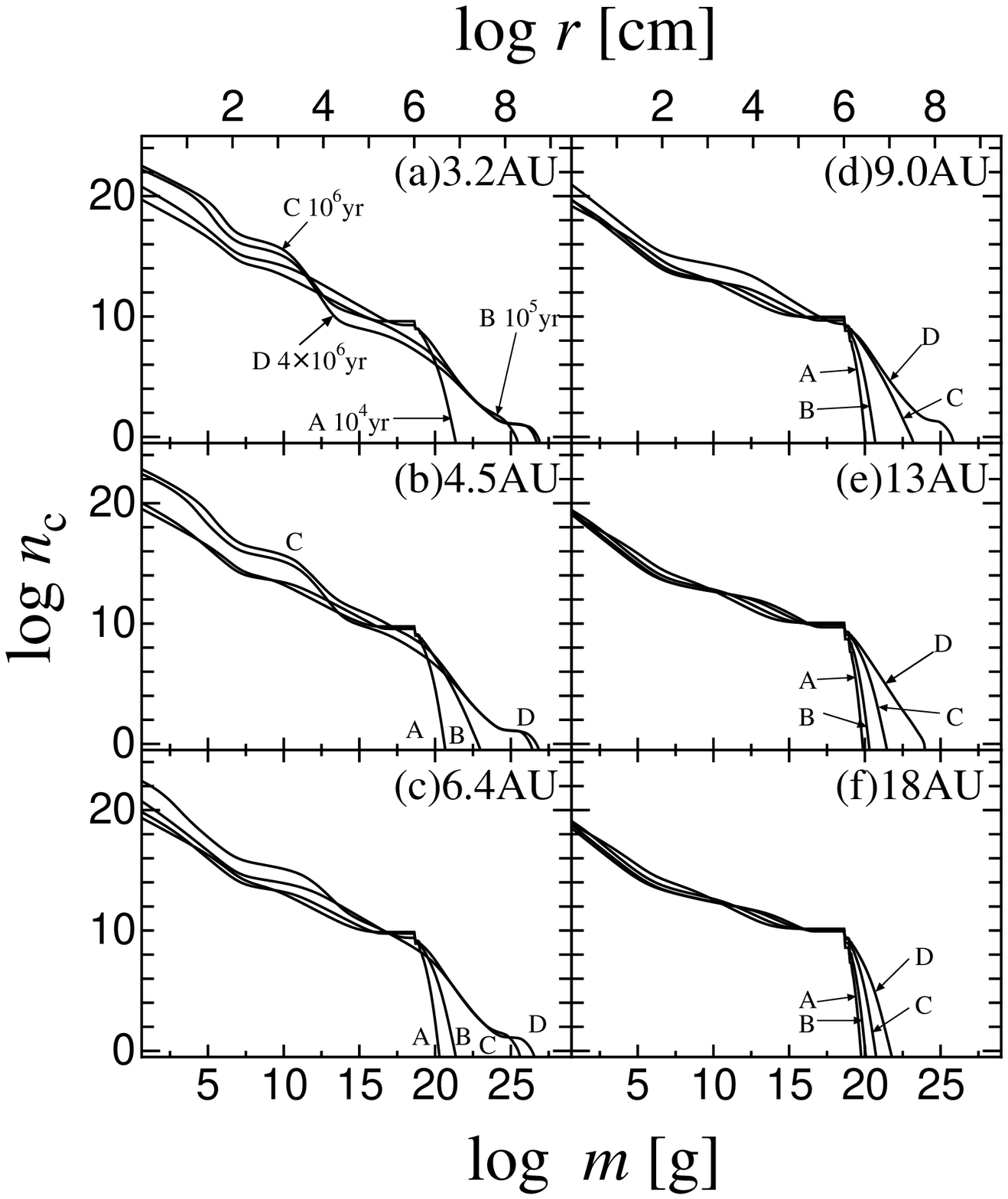} \figcaption{
\label{fig:mass_ev}}
\end{figure}

Fig.\ \ref{fig:mass_ev} --- Same as Fig.~\ref{fig:cumnum_nofrag}, but
with fragmentation.

\begin{figure}[htbp]
\epsscale{0.9} \plotone{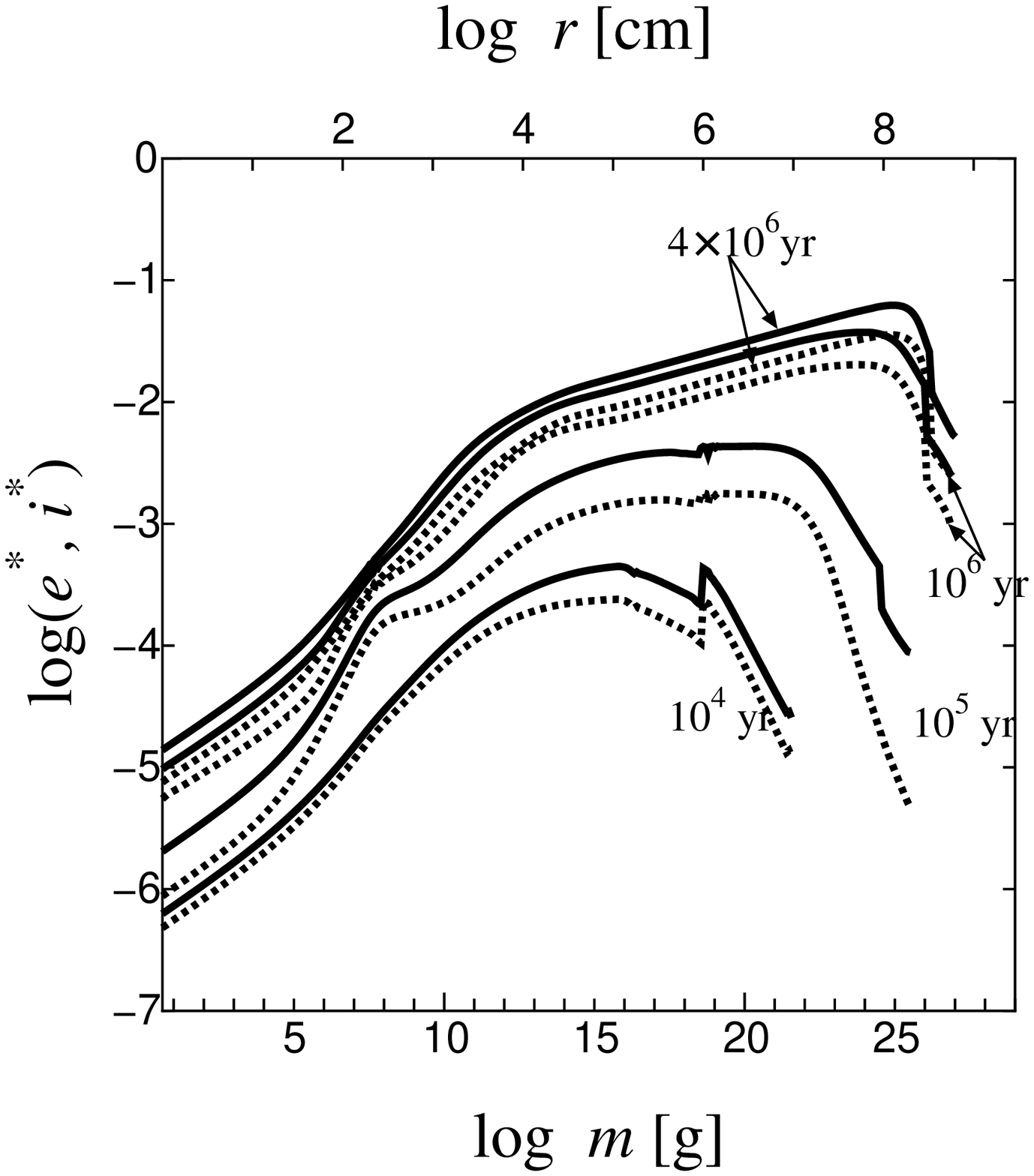} \figcaption{
\label{fig:vel_ei}}
\end{figure}

Fig.\ \ref{fig:vel_ei} --- Same as Fig.~\ref{fig:vel_nofrag}, but with
fragmentation.

\begin{figure}[htbp]
\epsscale{0.9} \plotone{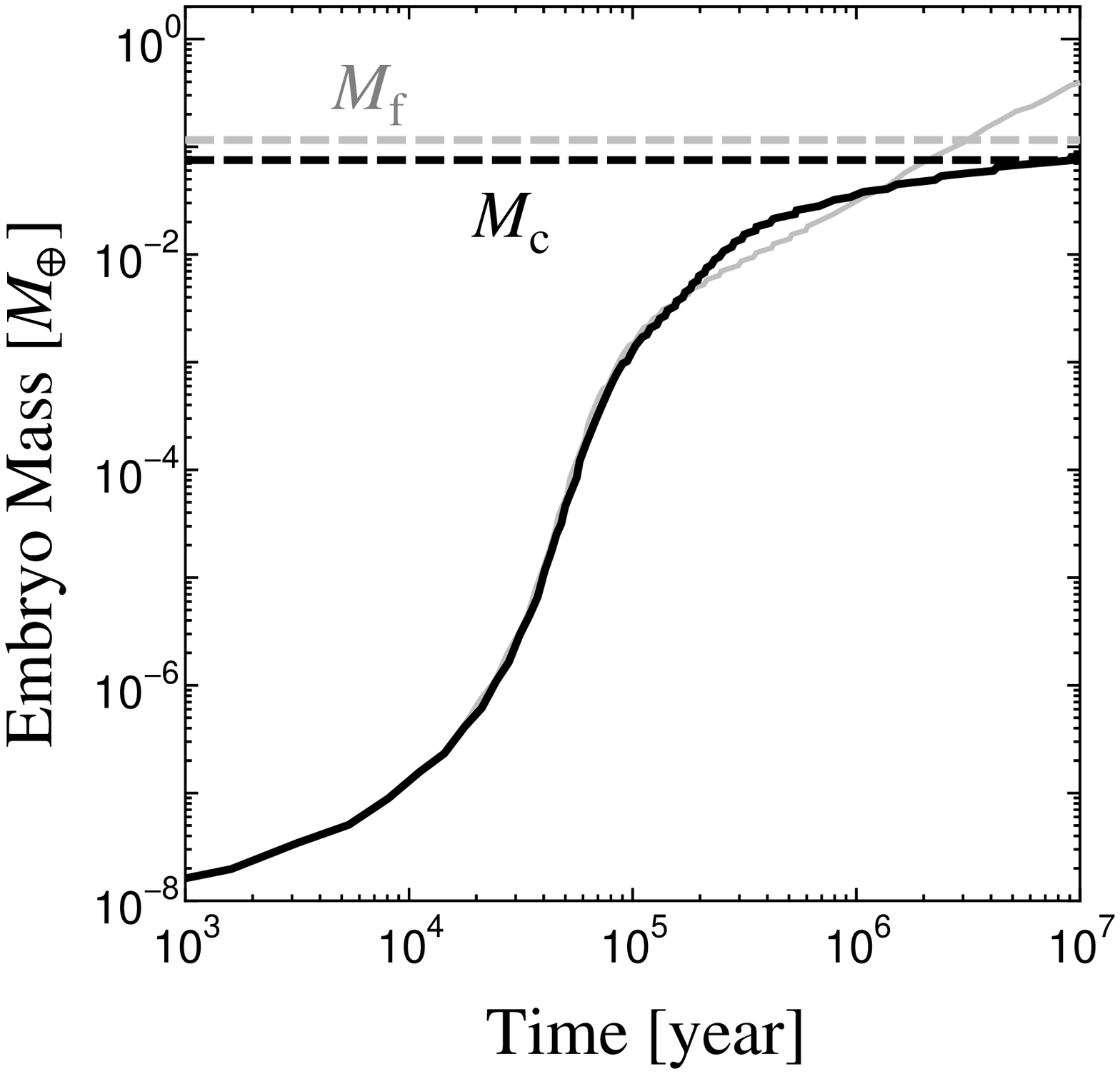} \figcaption{
\label{fig:emb_growth}}
\end{figure}

Fig.\ \ref{fig:emb_growth} --- Evolution of the embryo mass at 3.2\,AU
with fragmentation (black solid line) and without fragmentation (gray
solid line).  Dashed lines indicate $M_{\rm c}$ (Eq. \ref{eq:m_frag};
black) and $M_{\rm f}$ (Eq.~\ref{eq:mpla_frag}; gray).

\begin{figure}[htbp]
\epsscale{0.9} \plotone{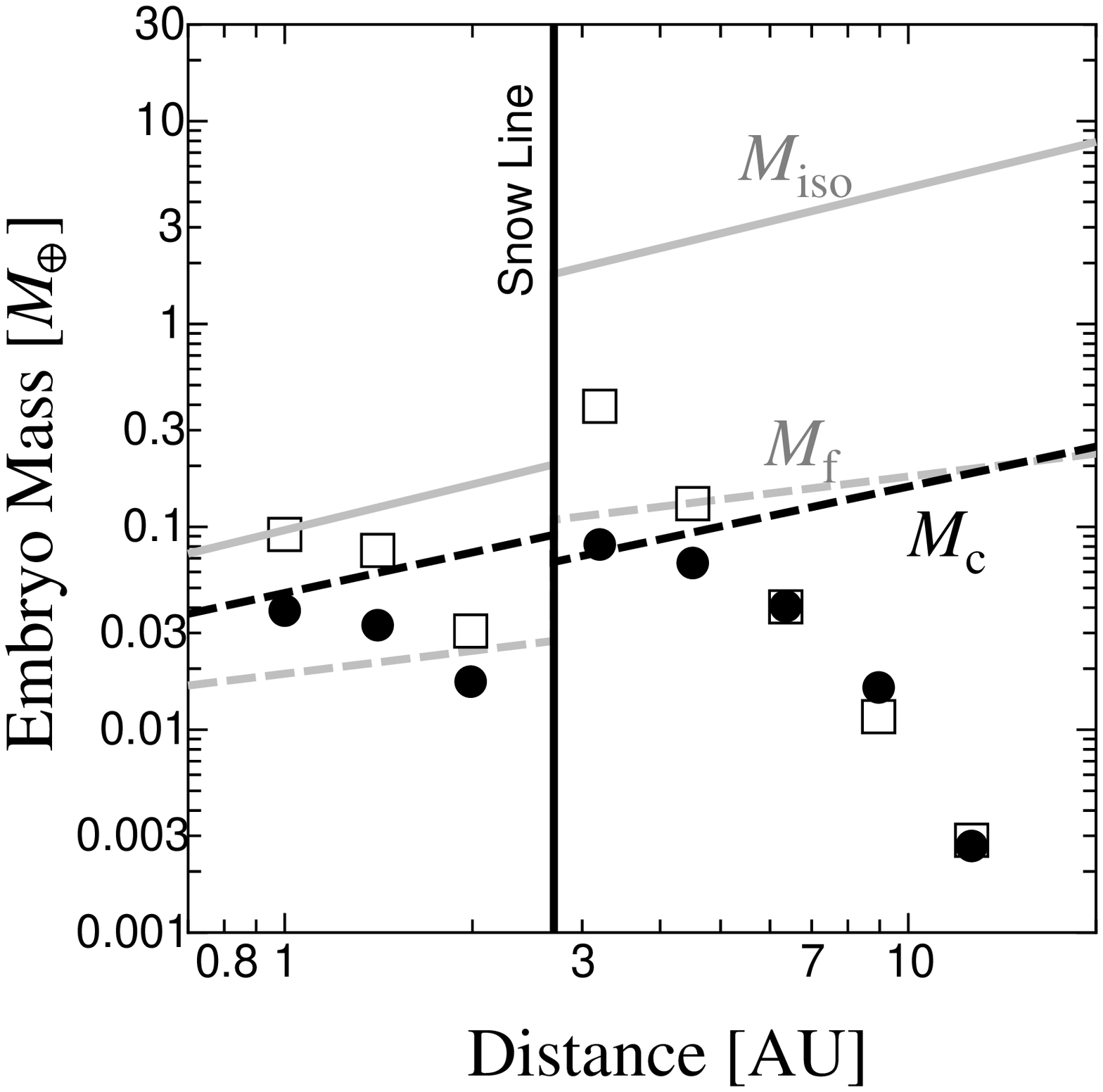} \figcaption{
\label{fig:final_mass}}
\end{figure}

Fig.\ \ref{fig:final_mass} --- The embryo mass at $10^7$ years for
initial planetesimal mass of $4.2\times 10^{18}$\,g with fragmentation
(filled circles) and without fragmentation (open squares).  Dashed lines
indicate $M_{\rm c}$ (Eq.~\ref{eq:m_frag}; black) and $M_{\rm f}$
(Eq.~\ref{eq:ratio_sigma}; grey).  Gray solid line shows $M_{\rm iso}$
(Eq.~\ref{eq:M_iso}) for comparison.  The vertical line represents the
snow line $a=a_{\rm ice} = 2.7\,$AU.

\begin{figure}[htbp]
\epsscale{0.9} \plotone{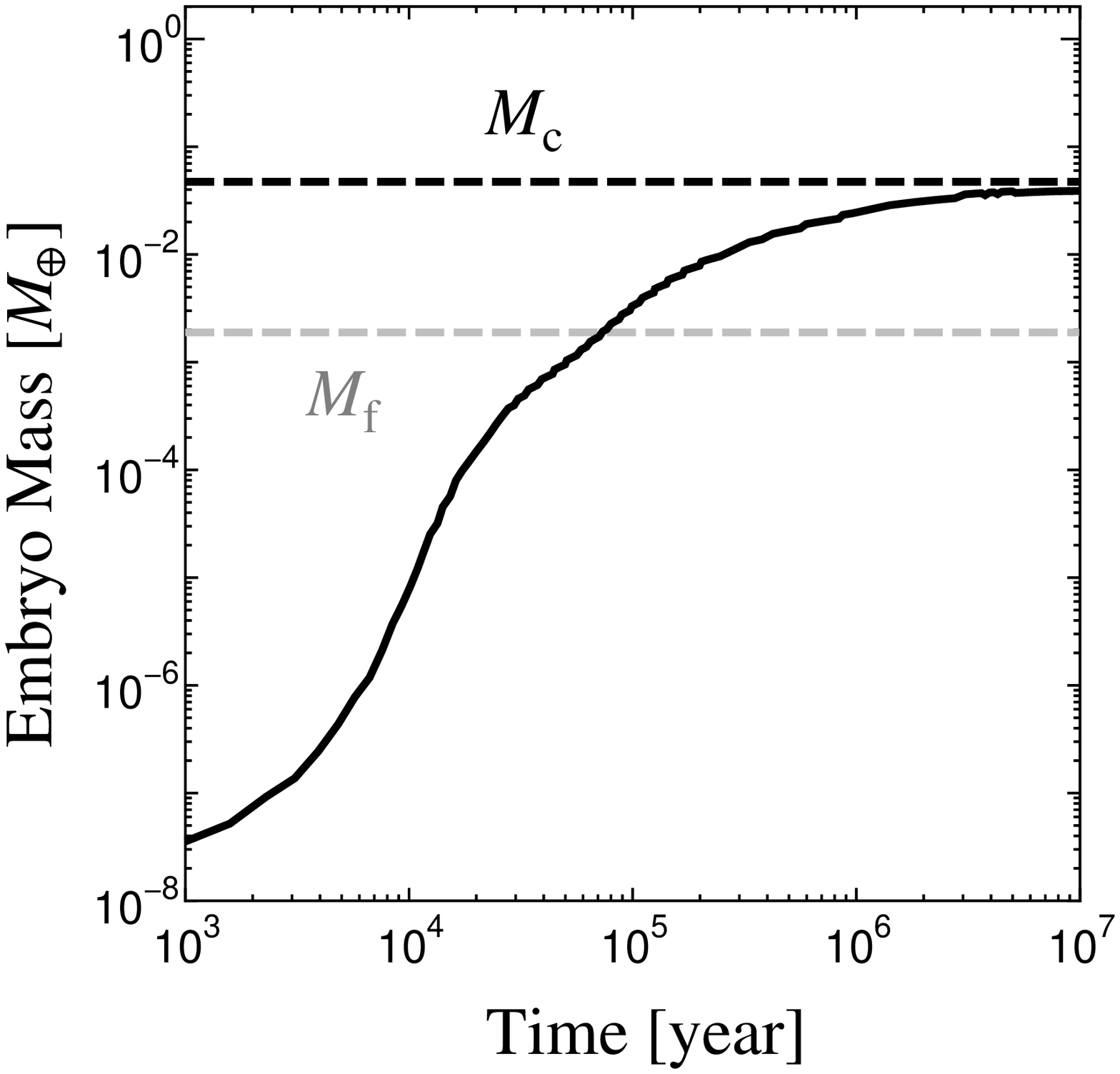} \figcaption{
\label{fig:emb_growth_at1au}}
\end{figure}

Fig.\ \ref{fig:emb_growth_at1au} --- 
Same as Fig.~\ref{fig:emb_growth}, but at 1\,AU. 

\begin{figure}[htbp]
\epsscale{0.9} \plotone{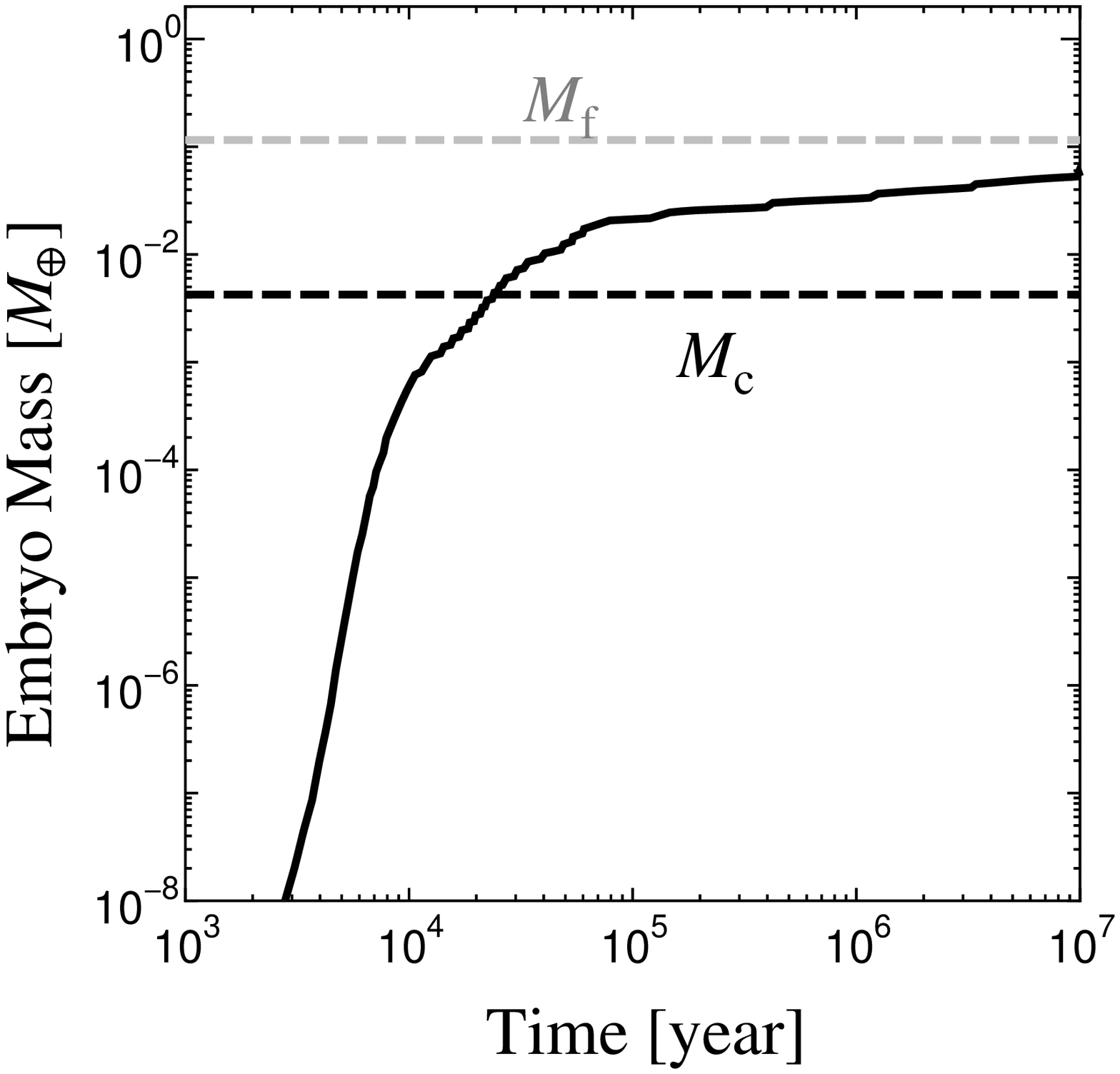} \figcaption{
\label{fig:emb_growth_r0_1}}
\end{figure}

Fig.\ \ref{fig:emb_growth_r0_1} --- Same as Fig.~\ref{fig:emb_growth},
but for $m_0 = 4.2\times 10^{15}$\,g ($r_0 = 1$\,km).

\begin{figure}
 \epsscale{0.9} \plotone{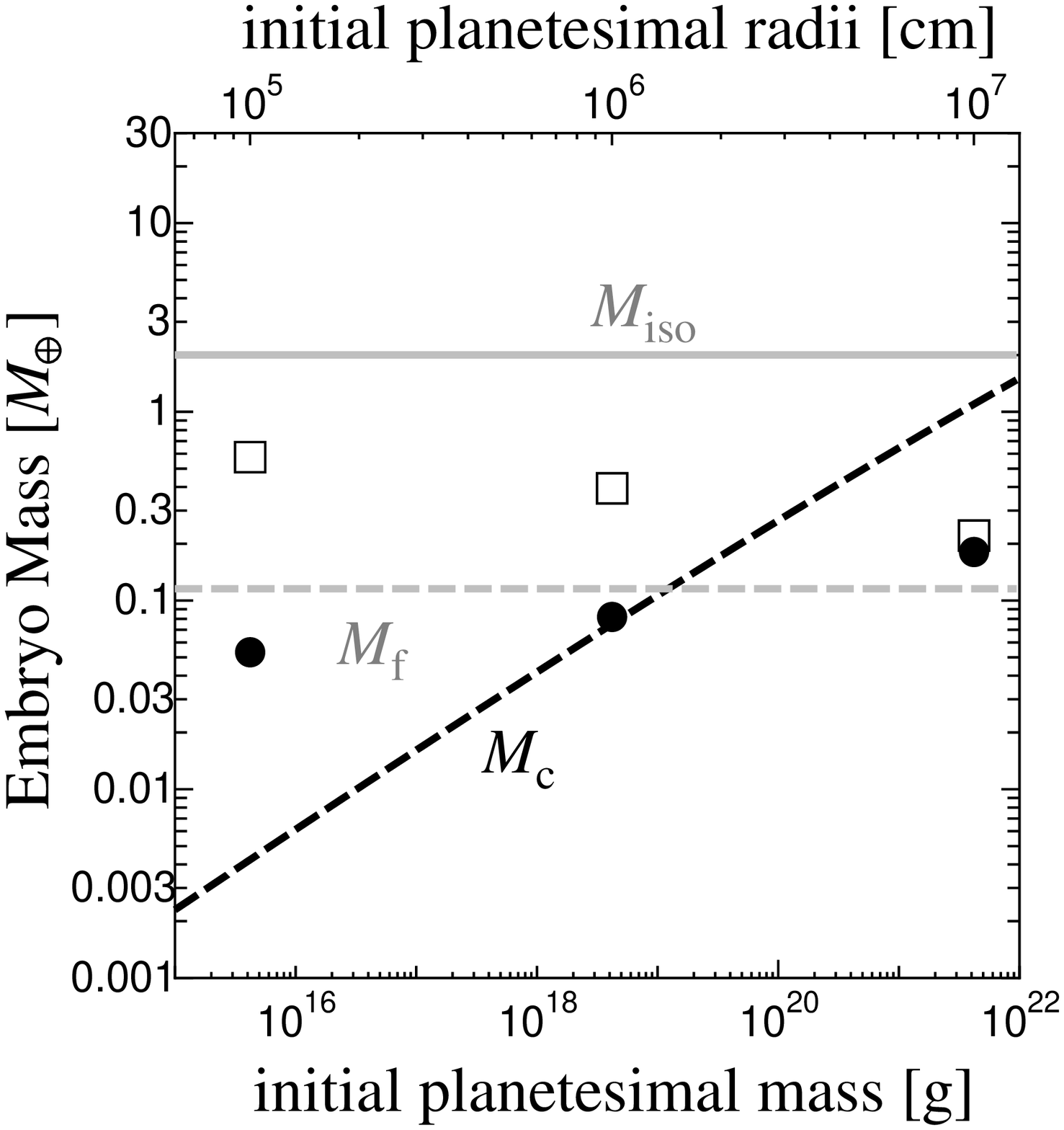} \figcaption{
\label{fig:fm_dep_r0}
}
\end{figure}

Fig.\ \ref{fig:fm_dep_r0} --- The embryo mass at 3.2\,AU at $10^7$ years
for the minimum-mass solar nebula model ($\Sigma_{\rm 1} = 7\,{\rm
g/cm}^2$ and $q=3/2$) as a function of the initial planetesimal mass.
The symbols and lines are same as Fig.~\ref{fig:final_mass}.

\begin{figure}
 \epsscale{0.9} \plotone{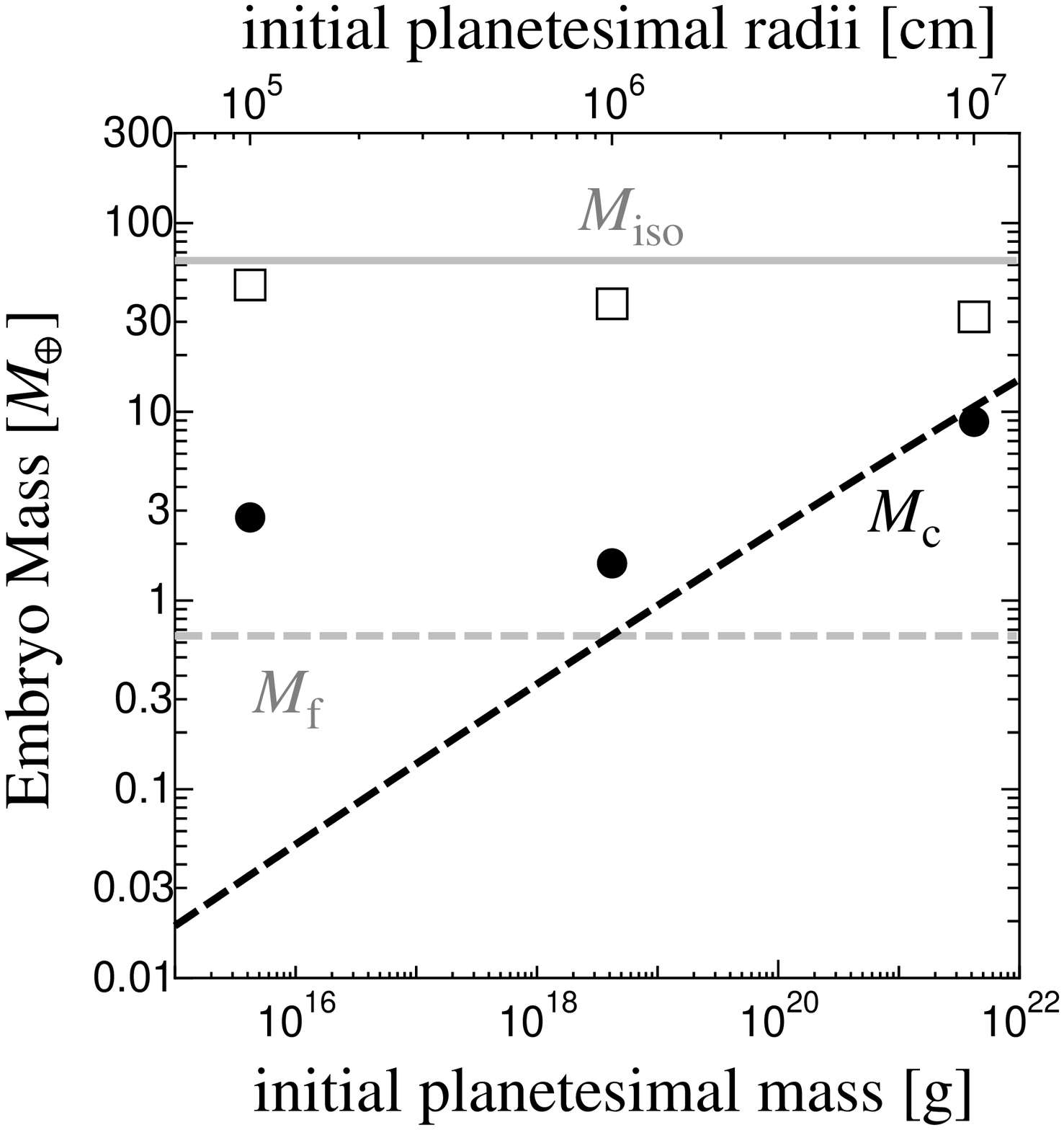} \figcaption{
\label{fig:fm_dep_r0_gamma10}
}
\end{figure}

Fig.\ \ref{fig:fm_dep_r0_gamma10} --- Same as Fig.~\ref{fig:fm_dep_r0},
but for $\Sigma_{1} = 70 \, {\rm g/cm}^2$.

\begin{figure}[htbp]
\epsscale{0.9} \plotone{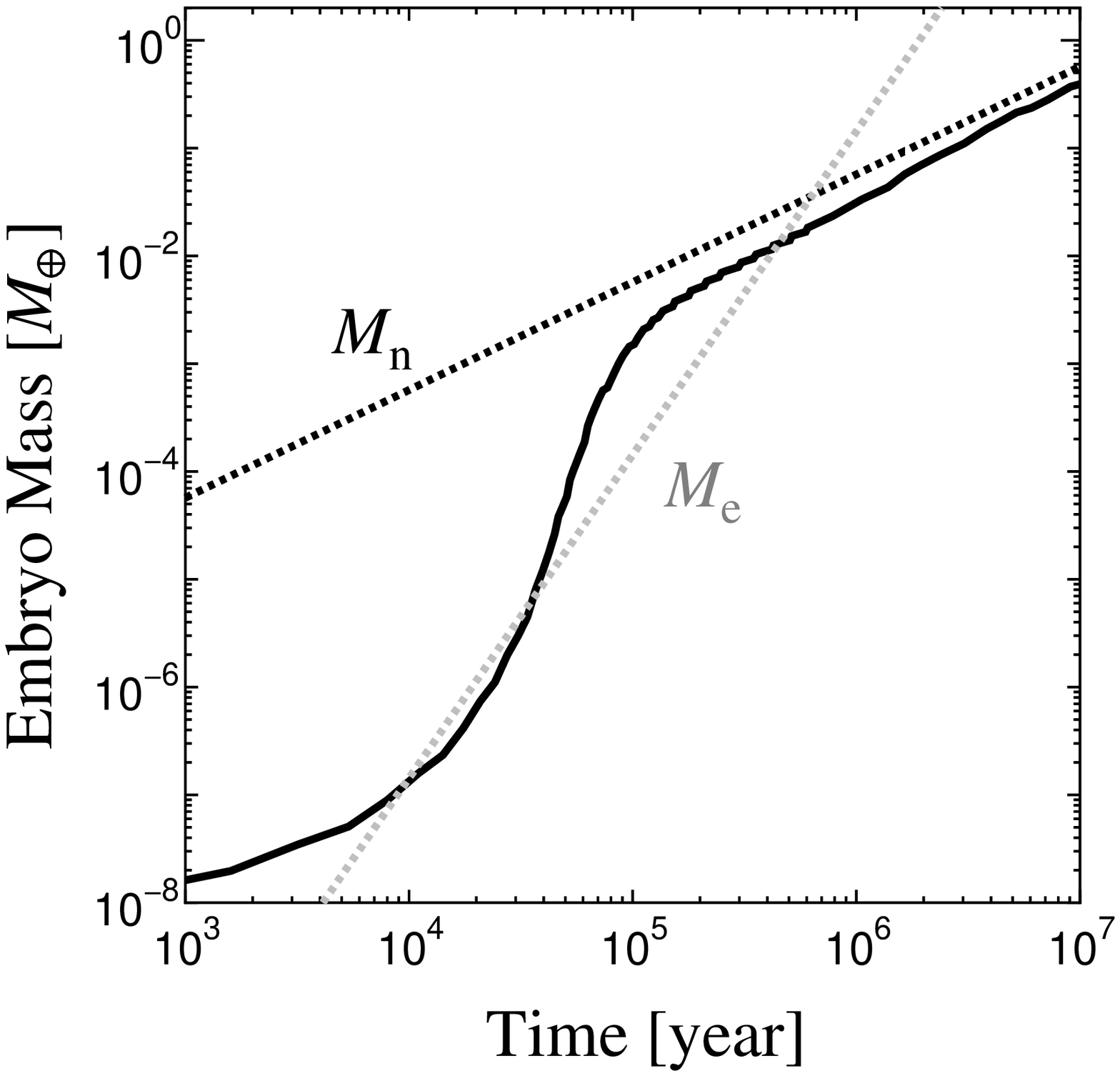} \figcaption{
\label{fig:emb_growth_nofrag}}
\end{figure}

Fig.\ \ref{fig:emb_growth_nofrag} --- 
Evolution of the embryo mass at 3.2\,AU, neglecting fragmentation. 
The dashed lines indicate $M_{\rm n}$ (Eq.~\ref{eq:m_nogas}; black) and
 $M_{\rm e}$ (Eq.~\ref{eq:m_gas}; gray).

\newpage 

\end{document}